\pdfoutput=1
\documentclass[prd,preprint,floats,floatfix,aps,nofootinbib,preprintnumbers]{revtex4}
\usepackage{graphics,graphicx,psfrag}
\usepackage{amsmath,amssymb}
\usepackage[sort&compress]{natbib}
\usepackage{bm}	
\usepackage{verbatim}
\usepackage{multirow}
\usepackage{subfigure}
\usepackage{ulem}
\usepackage{color}
\usepackage[linktocpage=true,bookmarksnumbered=true,colorlinks=true,plainpages,linkcolor=blue,citecolor=red]{hyperref}
\usepackage[utf8]{inputenc}  

\begin{document}
\preprint{
{\vbox {
\hbox{\bf MSUHEP-160719}
\hbox{\today}
}}}
\vspace*{2cm}

\title{Simplified Limits \\ on Resonances at the LHC}
\vspace*{0.25in}   
\author{R. Sekhar Chivukula$^1$}
\email{sekhar@msu.edu}
\author{Pawin Ittisamai$^2$}
\email{ittisama@msu.edu}
\author{Kirtimaan Mohan$^1$}
\email{kamohan@pa.msu.edu}
\author{Elizabeth H. Simmons$^1$}
\email{esimmons@msu.edu}
\affiliation{\vspace*{0.1in}
$^1$ Department of Physics and Astronomy\\
Michigan State University, East Lansing U.S.A.\\
$^2$ Department of Physics, Faculty of Science\\
Chulalongkorn University, Bangkok 10330, Thailand
}

\begin{abstract}
\vspace{0.5cm}
\noindent

In the earliest stages of evaluating new collider data, especially if a small excess may be present, it would be useful to have a method for comparing the data with entire classes of models, to get an immediate sense of which classes could conceivably be relevant.  In this paper, we propose a method that applies when the new physics invoked to explain the excess corresponds to the production and decay of a single, relatively narrow, $s$-channel resonance. A simplifed model of the resonance allows us to convert an estimated signal cross section into general bounds on the product of the branching ratios corresponding to the dominant  production and decay modes. This quickly reveals whether a given class of models could possibly produce a signal of the required size at the LHC.  Our work sets up a general framework, outlines how it operates for resonances with different numbers of production and decay modes, and analyzes cases of current experimental interest, including resonances decaying to dibosons, diphotons, dileptons, or dijets.  If the LHC experiments were to report their searches for new resonances beyond the standard model in the simplified limits variable $\zeta$ defined in this paper, that would make it far easier to avoid blind alleys and home in on the most likely candidate models to explain any observed excesses.

\end{abstract}

\maketitle

\section{Introduction}

Reams of data are flowing from LHC-13.  Some will be used to explore $2\to 2$ scattering processes where a narrow resonance arising from physics Beyond the Standard Model (BSM) is produced in the $s$-channel and immediately decays to visible final state particles.  At present, data is generally compared with theory by showing how the predictions of a  benchmark model with specific parameter choices compare to the observed limits on the cross-section ($\sigma$) times branching fraction ($BR$) for the process as a function of the resonance mass. A given experimental paper reporting new upper limits on $\sigma\cdot BR$ provides comparisons with just a handful of specific models. However, in the earliest stages of evaluating new data, especially when a small excess may be present, it would be far more useful to compare the data with entire classes of models, to get an immediate sense of whether a given class could conceivably be relevant.  In this paper, we introduce method for doing so.

At present most ``model-independent" analyses reported in searches for narrow BSM resonances are cast as a plot of the experimental upper limit on $\sigma \cdot BR$ plotted as a function of the mass of the new resonance. A set of theoretical prediction curves are overlaid on the data.  Generally each theory curve corresponds to a different choice of spin, electric charge, weak charge, and color charge for the new resonance; in that very general sense, the set of curves might be thought to span the theoretical possibilities.  But in reality, for a given choice of spin and charges, there will be multiple detailed theoretical realizations corresponding to very different strengths and chiralities of the resonance's couplings to the initial partons through which it is produced and to the final states into which it decays.  The single realization of a $W'$ or coloron shown in the analyses generally corresponds to a long-familiar example from the literature that is convenient to use because it has already been coded into PYTHIA \cite{Sjostrand:2014zea} or similar analysis tools.  Some such examples (like the leptophobic Z' boson) have no actual realization in any self-consistent models, but are used because they tend to have relatively large production rates.  

In contrast, we propose that reporting the results of collider searches for BSM resonances in terms of a different set of variables would make it possible to immediately discern whether {\it an entire class of resonances with particular dominant production modes and/or decay patterns} (e.g., a spin-zero state produced through gluon fusion and decaying to diphotons) could conceivably be responsible for a given deviation in cross-section data relative to standard model predictions.  When the answer is ``no", one need waste no further time proposing models based on that type of new resonance as an explanation for the excess.  When the answer is ``yes," one also obtains information on the range of masses and branching fractions a model would need to provide for the state in order for it to be compatable with the data; again, this could guide model-building into profitable directions.

This work builds off of our previous results on identifying the color \cite{Atre:2013mja,Chivukula:2014npa} and spin \cite{Chivukula:2014pma} properties of new resonances decaying to dijet final states.  In those papers, we noted how the color and spin of different resonances impacted the state's width, relative to fixed values of the production cross-section and mass.  This was encapsulated in the dimensionless color discriminant variable $D_{col} \equiv \sigma M^3 / \Gamma$.  Here we extend these ideas to a much wider variety of final states and to situations in which a resonance has not been measured, but rather a small deviation possibly indicative of a resonance has been observed.  

Evidence for or observation of an excess would generally be reported within a specific channel or set of a few channels.  Often, the BSM possibilities invoked to explain the excess correspond to the production
and decay of a single, relatively narrow, s-channel resonance. In this context, a simplifed model of the resonance allows us to convert any estimated signal cross section into general constraints on the properties of the resonance. More specifically, if resonance production occurs dominantly through a single process, we can obtain model-independent upper bounds on the product of the branching ratios corresponding to production and decay for that process. This can make it immediately clear whether a given class of models could possibly produce a signal of the required size at the LHC.  As we shall detail below, one can readily extend this to situations with more than one production or decay channel.

Other previous work in the literature is also relevant here. For example, EHLQ \cite{Eichten:1984eu} used parton luminosities to assist in assessing the potential {\it reach} of proposed new colliders; here, in contrast, we assess the ability of a specific resonance to explain a potential {\it signal} at an existing collider. Carena {\it et al.}  \cite{Carena:2004xs} classified $Z'$ bosons according to the BR to leptons and average couplings to quarks in order to compare multiple models with Tevatron data simultaneously; this was limited to $Z'$ bosons, was more model-dependent in its approach, and was aimed at determining discovery reach. Dobrescu \& Yu \cite{Dobrescu:2013coa} presented discovery limits on dijet resonances in a coupling vs. mass plane to facilitate comparison of results from different colliders; while our method could be used at a variety of colliders, we will give examples for the LHC.  Our work is focused on establishing a method for understanding the implications of current exclusion curves (not establishing discovery reaches) for entire classes of models; we find that using branching ratios, rather than couplings, is more effective in this context.\footnote{A more recent example with similarity to our approach is given in \cite{Franceschini:2015kwy}, in an analysis of potential sources for a potential diphoton signal \cite{ATLAS-Diphoton,Moriond-ATLAS,ATLAS-CONF-2016-018,CMS:2015cwa,CMS:2015dxe,Moriond-CMS,CMS-PAS-EXO-16-018} at the LHC.}

The next section sets up our general framework for the case of a narrow resonance and notes how the upper bounds on products of branching ratios are impacted by whether the initial partons are identical and whether the initial and final states differ from one another.  It then sketches how the upper bounds work for several cases with different numbers of resonance production and decay modes. It closes by introducing a new dimensionless variable, $\zeta$, that is related to the product of branching ratios but further simplifies analyses by reducing the  impact of the resonance's width.  Section 3 analyzes a number of cases of current experimental interest, including resonances decaying to dibosons, diphotons, dileptons, or dijets. Comparisons between observed and expected limits set by recent data and the upper limits on $\zeta$ yield a straightforward way of concluding whether a given class of resonance could explain a particular signal; extending the comparison to predictions of a specific model within a promising class then readily indicate whether that model is a viable candidate.  Section 4 discusses our results and future directions. Some underlying technical details of our calculations are summarized in the Appendices.  We suggest that if the LHC experiments were to report their searches for new resonances beyond the standard model in the simplified limits variable $\zeta$ defined in this paper, that would make it far easier to avoid blind alleys and home in on the most likely candidate models to explain any observed excesses.

\section{Narrow Resonances}

We aim to establish a framework for discussing the broad implications of experimental exclusion curves, which generally are couched in terms of specific sets of production and decay channels.  We will start by writing the cross-sections in terms of the branching ratios of the resonance to relevant final states.

The tree-level partonic production cross-section for an arbitrary 
$s$-channel resonance $R$ produced by collisions of particular initial state partons $i,j$ and decaying to  a single final state $x,y$ at the LHC can be written \cite{Harris:2011bh,Agashe:2014kda}
\begin{equation}
\hat{\sigma}_{ij\to R\to xy}(\hat{s}) = 16 \pi (1 + \delta_{ij}) \cdot {\cal N} \cdot
\frac{\Gamma(R\to i+j) \cdot \Gamma(R\to x+y)}
{(\hat{s}-m^2_R)^2 + m^2_R \Gamma^2_R} ~,
\end{equation}
where ${\cal N}$ is a ratio of spin and color counting factors\footnote{We note here that ${\cal N}$ depends on the color and spin properties of the incoming partons $i,j$. We will neglect this in what follows, assuming that this factor is the same for all relevant production modes in a given situation -- see discussion at the end of subsection \ref{subsec:general-case}. In fact, this assumption is valid in the great majority of cases.}
\begin{equation}
{\cal N} = \frac{N_{S_R}}{N_{S_i} N_{S_j}} \cdot
\frac{C_R}{C_i C_j},
\label{eq:N}
\end{equation}
where $N_S$ and $C$ count the number of spin- and color-states for initial state partons $i$ and $j$ and for the resonance $R$. In the narrow-width approximation, one can simplify this further, using the expression\footnote{In detail, resonance limits derived from observations will depend on whether $\Gamma_R/m_R$ lies below the experimental resolution for the invariant mass of the final state particles. As discussed in Appendix \ref{app:width}, however, these effects are expected to be only of order a factor of two -- meaning they are not relevant for the preliminary investigations envisioned here.}  
\begin{equation}
\frac{1}
{(\hat{s}-m^2_R)^2 + m^2_R \Gamma^2_R}
\approx \frac{\pi}{m_R \Gamma_R} \delta(\hat{s} - m^2_R)~.
\end{equation}

\noindent Integrating over parton densities, and summing over incoming partons and over the outgoing partons which produce experimentally indistinguishable final states (e.g., over final state light quarks $q\bar{q}$, with $q=u,d,s$ that produce
untagged two-jet final states), we then find the tree-level
hadronic cross section to be
\begin{align}
\sigma^{XY}_R &\ \equiv \sigma_R \times BR(R \to X + Y) = 16\pi^2 \cdot {\cal N} \cdot \frac{ \Gamma_R}{m_R} \times  \nonumber \\
& 
\left( \sum_{ij} (1 + \delta_{ij}) BR(R\to i+j) \left[\frac{1}{s} \frac{d L^{ij}}{d\tau}\right]_{\tau = \frac{m^2_R}{s}}\right) \cdot \left(\sum_{xy\, \in\, XY} BR(R\to x+y)\right)~. 
\label{eq:cross-section}
\end{align}
Here ${d L^{ij}}/{d\tau}$ corresponds to the luminosity function for the $ij$ combination of partons\footnote{
In particular,
	\begin{equation}
	\left[ \frac{d{L}^{ij}}{d\tau}\right] \equiv 
	\frac{1}{1 + \delta_{ij}} \int_{\tau}^{1} \frac{dx}{x}
			\left[ f_i\left(x, \mu_F^2\right) f_j\left( \frac{\tau}{x}, \mu_F^2 \right) +
			f_{j}\left(x, \mu_F^2\right) f_i\left( \frac{\tau}{x}, \mu_F^2 \right) \right]  \,,
	\label{eq:lumi-fun}
	\end{equation}
	where here, for the purposes of illustration, we calculate these parton luminosities using the {\tt CTEQ6L1}~\cite{Pumplin:2002vw} parton density functions, setting the factorization scale $\mu_F^2= m^2_R$. More details are given in Appendix \ref{app:parton-lum}.}, and $X\, Y$ label the
	set of experimentally indistinguishable final states.
	
This way of writing the cross-section lends itself well to judging which classes of models are capable of producing a given observable excess.  We will now walk through a variety of situations from the simplest, with resonances only produced and decaying in one way, through more complicated situations involving multiple production and decay modes.  In Section \ref{sec:applications}, we will treat specific instances of these scenarios in more detail.

It is important to note that while we have been calculating total cross-sections, some experimental results are given as limits on the total cross-section and others as limits on the cross-section times the acceptance due to kinematic cuts. Where we have encountered the latter, we have used simulations performed with {\tt MadGraphMC@NLO}~\cite{Alwall:2014hca} to evaluate the acceptance.\footnote{The acceptance due to kinematic cuts depends on the angular distribution of the final states, which in turn depends on the spin of the particles involved in the process. In cases with multiple production and decay modes, the acceptance therefore can change depending on the spins of the initial and final states \cite{Jacob:1959at,Haber:1994pe}. In particular, if there are multiple production modes with substantially different acceptances, one would have to consider these modes seperately. Note however, this does not affect any of the examples we have considered here. }

\subsection{Simplest Case: one production and one decay mode}

Let us first consider the simplest possible case in which only one set of initial $(i,j)$ and final $(x,y)$ states is relevant for production and decay of a new resonance $R$. 
A concrete example would be production of an up-flavored  excited quark: it will essentially be produced only through $ug$ fusion and decay predominantly back to $ug$, modulo contributions from modes accessible only via very small mixing angles (lest there be large flavor-changing neutral currents). Production or decay through $u\gamma$ (or $dW$) would be suppressed by the smaller $u\gamma$ ($dW$) coupling and the relatively small $\gamma$ ($W$) parton luminosity.

We can write down the signal cross-section for pp-collisions as follows (here $XY$ reduces to $xy$ because there is only one decay mode),
	\begin{equation}
	\sigma^{xy}_R=\sigma_R \times BR(R \to x+y) = \int_{s_{min}}^{s_{max}}d\hat{s}\,
	\hat{\sigma}(\hat{s}) \cdot \left[ \frac{d L^{ij}}{d\hat{s}}\right]~,
	\label{eq:simplest1}
	\end{equation}
and hence, in the narrow-width approximation,
\begin{equation}
\sigma^{xy}_R = 16\pi^2 \cdot {\cal N} \cdot \frac{ \Gamma_R}{m_R}  \cdot
 (1 + \delta_{ij})BR(R\to i+j) \left[\frac{1}{s} \frac{d L^{ij}}{d\tau}\right]_{\tau = \frac{m^2_R}{s}}  \cdot BR(R\to x+y)~.
\label{eq:simplest2}
\end{equation}

This can be reframed as an expression for the product of branching ratios:
\begin{equation}
   BR(R\to i+j) (1 + \delta_{ij}) \cdot BR(R\to x+y)  =  \frac{\sigma^{xy}_R}{16 \pi^2  {\cal N} \frac{\Gamma_R}{m_R} \left[\frac{1}{s} \frac{d L^{ij}}{d\tau}\right]_{\tau = \frac{m^2_R}{s}}}~.
\label{eq:simplebound-lower}
\end{equation}
This equation essentially tells us that if an {\it arbitrary} $s$-channel resonance with a given value of $\Gamma_R/m_R$ produced from partons $ij$ is to produce a signal of a particular size, then the product of the resonance's branching ratios must attain a certain value. Significantly, this value depends only on the properties of the resonance and the partonic luminosity of the initial state partons.  It can therefore be used to distinguish among potential theoretical descriptions of any new resonance.

At the same time, since the sum of all branching ratios of a resonance equals one, we can set a theoretical upper bound on the value of the product of branching ratios discussed above. There are four possibilities. First, assume the incoming partons are not identical, so that $i\neq j$.  There are two sub-cases:
\begin{itemize}
\item If the initial and final states differ from one another ($ij \neq xy$, as in the process $gu \to R \to gt$),  we necessarily find that the LHS of Eqn. \ref{eq:simplebound-lower}, namely $BR(R\to i+j) (1 + \delta_{ij}) \cdot BR(R\to x+y)$ has a value $\le 1/4$. 
\item If the initial ($ij$) and final$(xy)$ states are the same, ($ij = xy$, as in $WZ \to R \to WZ$), then the LHS of Eqn. \ref{eq:simplebound-lower} is $\le 1$. 
\end{itemize}
In contrast, if we assume the incoming partons are identical ($i=j$) then $(1 + \delta_{ij}) = 2$.  This raises the upper bounds:
\begin{itemize}
\item If the initial and final states differ ($ij\neq xy$, as in $gg \to R \to \gamma\gamma$), then the LHS of Eqn. \ref{eq:simplebound-lower} is $\le 1/2$.
\item If they are the same ($ij = xy$, as in $\gamma\gamma \to R \to \gamma\gamma$) , then the LHS of Eqn. \ref{eq:simplebound-lower} is $\le 2$.
\end{itemize}

Experimental searches for a narrow resonances $R\to x+y$ are generally reported in terms of expected and observed upper bounds in the $\sigma^{xy}_R \equiv \sigma(pp \to R)\cdot  BR(R\to x+y)$ vs. $m_R$ plane. A potential narrow resonance appears initially (prior to a 5$\sigma$ discovery) as a deviation in which the observed limit is weaker than the expected limit. When such a deviation is seen, one immediately asks what kinds of resonances $R \to x+y$ could potentially explain this excess.  The tendency has been to make comparisons with very specific models.  

We suggest that a more general approach based on Eqn. \ref{eq:simplebound-lower} can be far more informative. Specifically, the value of the product of branching ratios required to achieve a given $\sigma_R$ can be plotted on the same plane for various choices of $ij$ and $R$ and compared with the upper bounds on that product of branching ratio (either 1/4, 1/2, 1, or 2) derived above.  One will immediately see which classes of resonances could potentially give rise to the observed deviation.  We will illustrate this in detail in Section \ref{sec:applications}.

\subsection{Nearly-Simplest Case: one production mode, multiple decay modes}

Here, at most one of the decay modes is available as a production mode because the other decay modes involve states with negligible parton distribution functions.   An example could be a colored scalar with significant couplings to $gg$ and $t\bar{t}$.  Because the top quark PDFs are so small, the scalar will be produced overwhelmingly via $gg$ fusion; but it may have significant branching fractions to both $gg$ and $t\bar{t}$ final states.

The single production mode is handled as above. The sum of the outgoing branching ratios now plays the role that the single decay branching ratio played earlier.  In particular, Eqn. \ref{eq:simplebound-lower} now takes the form
\begin{equation}
   BR(R\to i+j) (1 + \delta_{ij}) \cdot \sum_{xy\,\in\, XY} BR(R\to x+y)  =  \frac{\sigma^{XY}_R}{16 \pi^2  {\cal N} \frac{\Gamma_R}{m_R} \left[\frac{1}{s} \frac{d L^{ij}}{d\tau}\right]_{\tau = \frac{m^2_R}{s}}}~.
\label{eq:nearlybound-lower}
\end{equation}

Since all branching ratios are positive and the sum over all branching ratios is 1, we still have the same four possibilities for upper bounds on the combination of branching ratios as before. We therefore find
\begin{equation}
BR(R\to i+j) (1 + \delta_{ij}) \cdot \sum_{xy\,\in\, XY} BR(R\to x+y)  \le
\begin{cases}
1/4 & i\neq j,\, ij \neq xy\in XY \\
1 & i\neq j,\, ij = xy\in XY \\
1/2 & i=j,\, x=y,\, ij\neq xy\in XY \\
2 & i=j, x=y, ij = xy\in XY
\end{cases}
\label{eq:combined}
\end{equation}
Now one would use these upper bounds in combination with Eqn. \ref{eq:nearlybound-lower} as the basis of comparing theory with experiment. Applications will be discussed in Section \ref{sec:applications}.

\subsection{General Case: Multiple production and decay modes }
\label{subsec:general-case}

This situation is complicated by the fact that the branching ratio for each initial state ($ij$) is associated with the luminosity function for that particular pair of partons.  An important example is in the case of the production of a $Z'$ which can proceed through either $u\bar{u}$ or $d\bar{d}$ annhilation, two modes with comparable partonic luminosities. We will need to rewrite Eqn.~\ref{eq:cross-section} in order to relate theoretical upper limits on products of branching ratios to the value of the cross-section, resonance properties, and parton luminosities.

The sum over branching ratios times luminosities for incoming partons $ij$ in the lower line of Eqn.~\ref{eq:cross-section} may be usefully reframed by simultaneously multiplying and dividing it by a sum over just incoming parton branching ratios (now labeled as $i' j'$), specifically: $ \sum_{i'j'} (1+\delta_{i'j'}) BR(R\to i' + j')$ 
\begin{align}
\sum_{ij} & (1 + \delta_{ij}) BR(R\to i+j) \left[\frac{1}{s} \frac{d L^{ij}}{d\tau}\right]_{\tau = \frac{m^2_R}{s}} = \\
&\left[\sum_{ij} \omega_{ij} \left[\frac{1}{s} \frac{d L^{ij}}{d\tau}\right]_{\tau = \frac{m^2_R}{s}}\right] \cdot \left[\sum_{i'j'} (1 + \delta_{i'j'}) BR(R\to i'+j') \right] \nonumber 
\label{eq:rewritten}
\end{align}
where
\begin{equation}
\omega_{ij} \equiv \dfrac {(1 + \delta_{ij})  BR(R\to i+j)} {\sum_{i'j'} (1 + \delta_{i'j'}) BR(R\to i'+j')}~.
\end{equation}
The fraction $\omega_{ij}$ lies in the range $ 0 \le \omega_{ij} \le 1$ and by construction $\sum_{ij} \omega_{ij} = 1$.  
Essentially, $\omega_{ij}$ tells us the weighting of each set of parton luminosities $L^{ij}$.

Returning now to the expression for the cross-section in Eqn.~\ref{eq:cross-section}, we have\footnote{Again, as noted in footnote 1 above, we are assuming that all relevant production modes share the same value of the color and spin factor ${\cal N}$.}
\begin{align}
\sigma^{XY}_R & = 16\pi^2 \cdot {\cal N} \cdot \frac{ \Gamma_R}{m_R} \cdot 
\left[\sum_{ij} \omega_{ij} \left[\frac{1}{s} \frac{d L^{ij}}{d\tau}\right]_{\tau = \frac{m^2_R}{s}}\right] \cdot  \\
&  \cdot \left[\sum_{i'j'} (1 + \delta_{i'j'}) BR(R\to i'+j') \right]  \cdot \left(\sum_{xy\, \in\, XY} BR(R\to x+y)\right)~, \nonumber
\end{align}
which we can re-arrange to give an expression for the product of the sums of incoming and outgoing branching ratios:
\begin{align}
\left[\sum_{i'j'} (1 + \delta_{i'j'}) BR(R\to i'+j') \right]  &\cdot \left(\sum_{xy\, \in\, XY} BR(R\to x+y)\right) = 
\label{eq:gen-bran-init}\\
&\frac{\sigma^{XY}_R} { 16\pi^2 \cdot {\cal N} \cdot \frac{\Gamma_R}{m_R} \times 
\left[\sum_{ij} \omega_{ij} \left[\frac{1}{s} \frac{d L^{ij}}{d\tau}\right]_{\tau = \frac{m^2_R}{s}}\right]} ~, \nonumber
\end{align}
which generalizes Eqns. \ref{eq:simplebound-lower} and \ref{eq:nearlybound-lower}. The upper bound on the product of sums over branching ratios will be $1/4$, $1/2$, $1$ or $2$, depending on the identities of the incoming ($i'j'$) and outgoing ($x,y$) partons, in a straightforward generalization of Eqn. \ref{eq:combined}.

In closing this subsection, let us return to consider a limitation of the analysis that has been presented above. Namely, our analysis has implicitly assumed that all relevant production modes of a given resonance have the same color and spin properties -- {\it i.e.} that the value of ${\cal N}$ defined in Eq. \ref{eq:N} is the same for all relevant production modes. In general, this need not be the case: consider, for instance, a neutral scalar boson that couples to both gluons and photons. The total production cross section would include both a $gg$ luminosity factor with ${\cal N} = 1/256$, and a $\gamma \gamma$ one with ${\cal N}=1/4$. 
The formalism described above in Eqn. \ref{eq:gen-bran-init} would not work.
However, the corresponding luminosity functions for gluon and photon fusion differ by many orders of magnitude. In practice, therefore, one would consider $gg$ and $\gamma\gamma$ fusion separately (each as a case with one production mode and multiple decay modes); only if both production mechanisms turned out to be potentially relevant to the data would one need to undertake a more sophisticated analysis simultaneously involving both production modes. 

\subsection{Simplified Language}

Finally, we note that it is actually easier to make comparisons between data and theory if one re-arranges Eqn.~\ref{eq:gen-bran-init} (and analogously  Eqns.~\ref{eq:simplebound-lower} and \ref{eq:nearlybound-lower}) slightly so that the left-hand side includes the ratio of resonance width to mass. This enables us to define a useful dimensionless quantity which we will call $\zeta$:
\begin{align}
\zeta \equiv \left[\sum_{i'j'} (1 + \delta_{i'j'}) BR(R\to i'+j') \right]  &\cdot \left(\sum_{xy\, \in\, XY} BR(R\to x+y)\right) \cdot \frac{\Gamma_R}{m_R} = 
\label{eq:gen-bran-rat}
\\
&\frac{\sigma^{XY}_R} { 16\pi^2 \cdot {\cal N} \times 
\left[\sum_{ij} \omega_{ij} \left[\frac{1}{s} \frac{d L^{ij}}{d\tau}\right]_{\tau = \frac{m^2_R}{s}}\right]} ~. \nonumber
\end{align}
Because we are working in the narrow width approximation, and assuming that $\Gamma/M \leq 10\%$, the upper bounds on the products of branching ratios mentioned earlier may now be translated into upper limits on $\zeta$ that are a factor of ten smaller.  

In the examples below, we will first illustrate how to think about a few cases in terms of branching ratios, and then translate into making comparisons based on $\zeta$.  Subsequent examples will be explored in terms of $\zeta$ alone, because it is more versatile.

\section{Applications}
\label{sec:applications}

We will now apply the simplified limits technique to various situations of general theoretical and experimental interest.  These will include scalars decaying to diphotons, dijets, or $t\bar{t}$; a spin-1 state decaying to dibosons or dijets; a spin-2 state decaying to diphotons or dijets; a $W^\prime$ boson decaying to $WZ$ or dijets; and a $Z^\prime$ decaying to charged dileptons or dijets. We will organize this work according to the categorization above by number of production and decay modes.  

For instance, if faced by an apparent excess in a diboson final state, one might wonder whether any model of a $W^\prime$ decaying to dibosons could be the cause.  To answer that question, it would be important to  distinguish between (a) the case in which a strictly fermiophobic $W^\prime$ is produced by diboson fusion, (b) the case in which a $W^\prime$ couples sufficiently to dijets to be produced by quark/anti-quark annihilation, and (c) the variation of the latter in which the $W^\prime$ also has a significant decay fraction into dijets.  The first would be an instance of the simplest case where the single production and decay modes match; the second, the simplest case where the unique production and decay modes differ; the third, the nearly-simplest case where there are two decay modes of consequence. As we will see, the analysis and conclusions can be quite different in these cases, which would give crucial guidance for further model-building and phenomenological work.

A few technical details that are neglected in this section are discussed in the appendices. First, the width of the resonance can have an impact on the limits and we discuss this in Appendix~\ref{app:width}. Second, the calculation of the parton luminosities is discussed in Appendix~\ref{app:parton-lum}. 

\subsection{Examples of The Simplest Case: one production and one decay mode}

We will start with examples in which only one set of initial $(i,j)$ and final $(x,y)$ states is relevant for production and decay of a new resonance $R$.  In some cases, those initial and final states may be identical; in others, they will differ. 

\subsubsection{Fermiophobic $W^\prime$: $W_L^{\pm} Z_L \to W^{\prime} \to W_L^{\pm} Z_L$}
\label{subsec:fermiophobic-W-prime}

Our first example will be a charged spin-one color-neutral vector resonance -- a technirho or a $W^\prime$ -- that is primarily produced by vector boson fusion and primarily decays to $W_L^\pm Z_L$.  Because the initial and final states are identical (but the two incoming partons differ from one another), the signal cross section (in this simplified model) is determined entirely by $BR(R \to W_L Z_L)$, which cannot exceed 1.   

It is interesting to inquire whether such a resonance could have been responsible for the diboson excesses reported in the summer 2015 data from ATLAS and CMS (see, e.g., refs 1-14 of \cite{Brehmer:2015dan}). In the analysis reported in \cite{Aad:2015owa} based on hadronically-decaying dibosons, the most significant discrepancy in the $WZ$ channel from the background-only hypothesis occurs at an invariant mass of order 2 TeV; the local significance is $3.4\sigma$ and the global significance including the look-elsewhere effect in all three channels ($WZ, WW, ZZ$) is $2.5\sigma$.

\begin{figure}[btp]
	\centering
	\includegraphics[width=0.49\textwidth]{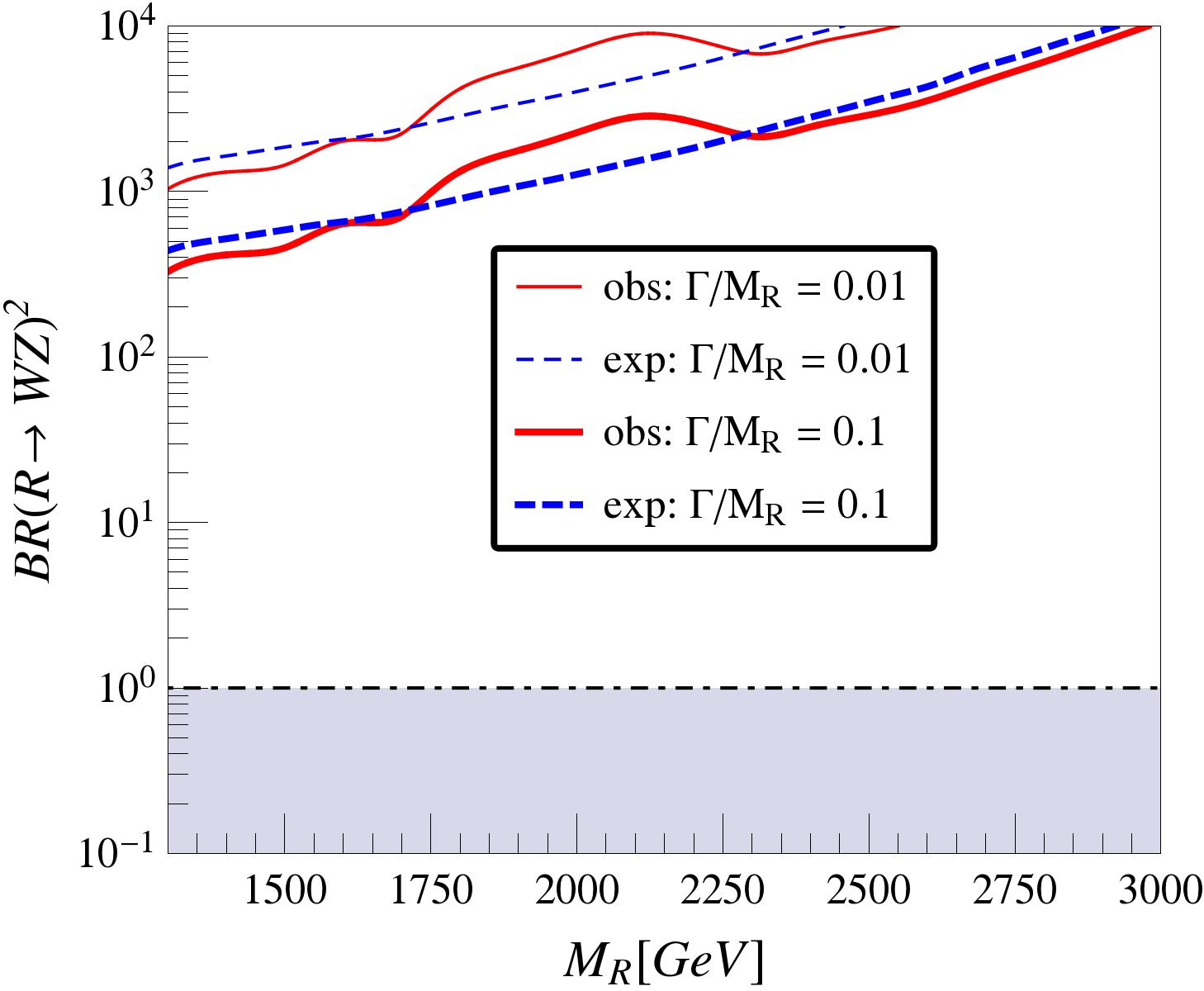}
	\includegraphics[width=0.49\textwidth]{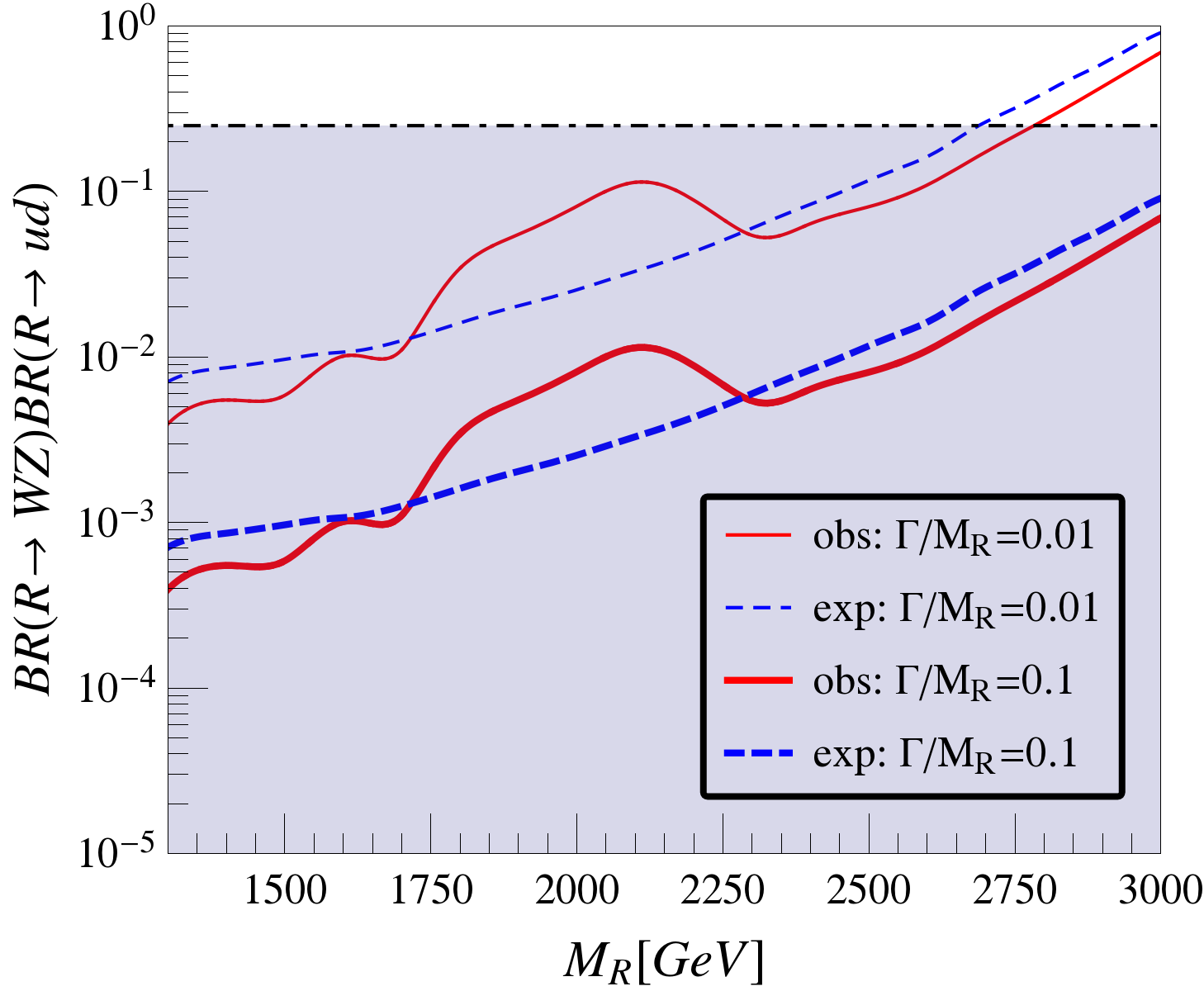}
	\caption{\small \baselineskip=3pt {\textbf{Left: }}The experimental ATLAS \cite{Aad:2015owa} upper limits (solid red curves) and expected limits (dashed blue curves) yield these upper bounds on the branching ratio product $[BR(R\to W_LZ_L)]^2$ assuming production of an s-channel resonance $R$ via vector boson fusion alone; results are shown for two values of $\Gamma/m_R$. As discussed in the text, since the apparent excess lies well outside the allowed (shaded) region, this scenario is disfavored. {\textbf{Right: }}The experimental ATLAS \cite{Aad:2015owa} upper limits (solid red curves) and expected limits (dashed blue curves)  yield these upper bounds on the branching ratio product $[BR(R\to u\bar{d} + d\bar{u})][BR(R\to W_LZ_L)]$ assuming production of an s-channel resonance $R$ via $u\bar{d} + d\bar{u}$ annihilation alone, shown for two values of $\Gamma/m_R$. As discussed in the text, since the apparent excess lies well within the allowed (shaded) region, this scenario was viable. }
	\label{fig:simplified-vbf2-ud2}
\end{figure}

In the left pane of ~\ref{fig:simplified-vbf2-ud2}, we have applied Eq. \ref{eq:simplebound-lower} to the observed and expected experimental upper limits \cite{Aad:2015owa} on the production cross-section for a resonance produced by $WZ$ fusion and decaying back to the same state.\footnote{We estimate the $WZ$ parton luminosities using the Effective W approximation~\cite{Dawson:1984gc,Dawson:1984gx,Kane:1984bb}, details of which can be found in Appendix~\ref{app:parton-lum}.} As this requires one to assume a specific value for the resonance's width/mass ratio, we show the results for $\Gamma_R / M_R = 1\%,\ 10\%$.   In the region around a resonance mass of 2 TeV, the observed upper bound is weaker than expected, indicating that an excess may be present.  From the plot, it is clear that the squared branching ratio $[BR(R \to WZ)]^2$ required to produce the excess production rate would be of order a few hundred to a few thousand. This greatly exceeds the maximum possible value of 1; allowed values of the squared branching ratio fall in the shaded region towards the bottom of the pane. Therefore, longitudinal vector boson fusion cannot be the dominant production mode for any $W^{\prime}$ resonance responsible for the observed possible diboson excess.  

\begin{figure}[btp]
	\centering
	\includegraphics[width=0.49\textwidth]{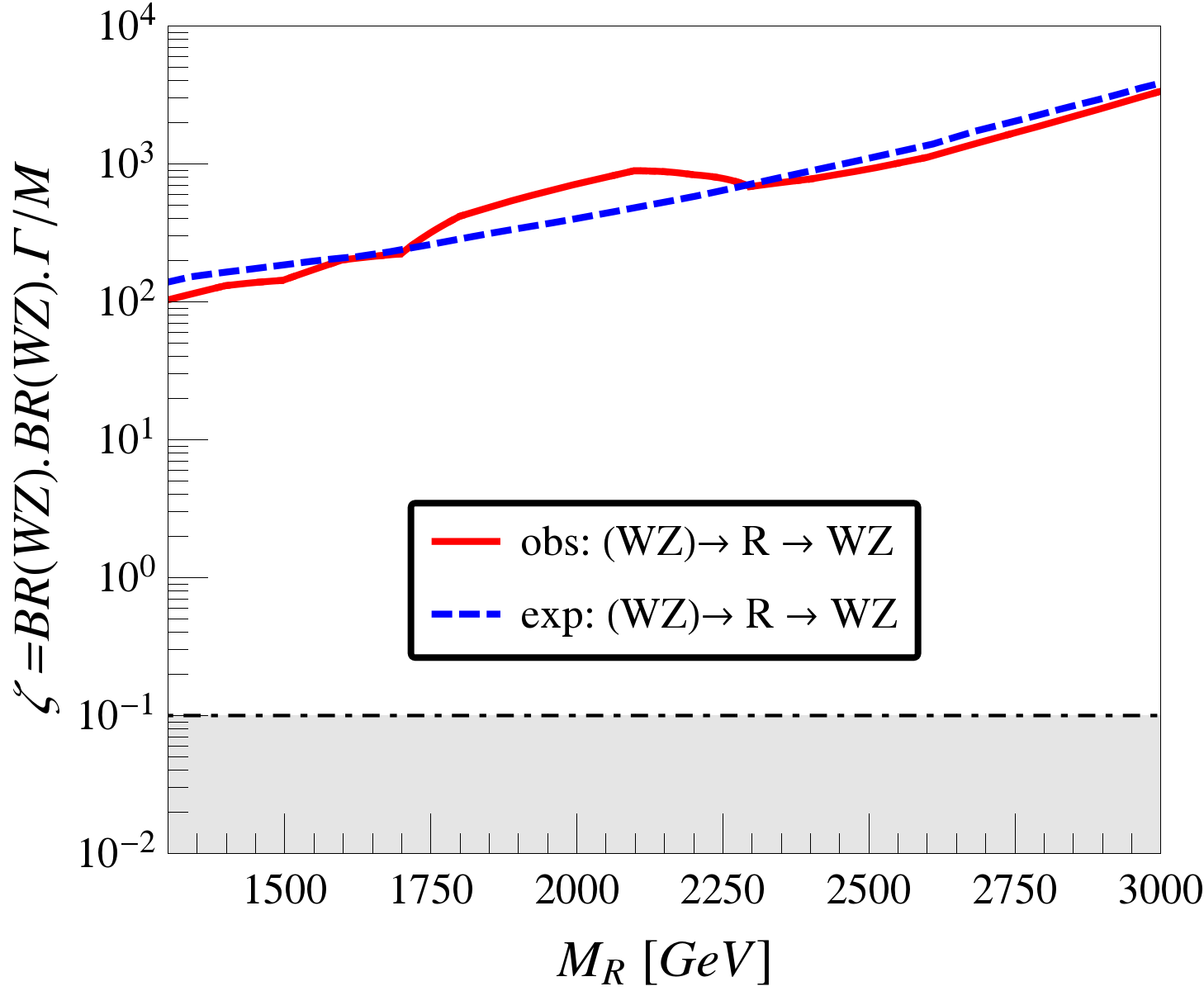}
	\includegraphics[width=0.49\textwidth]{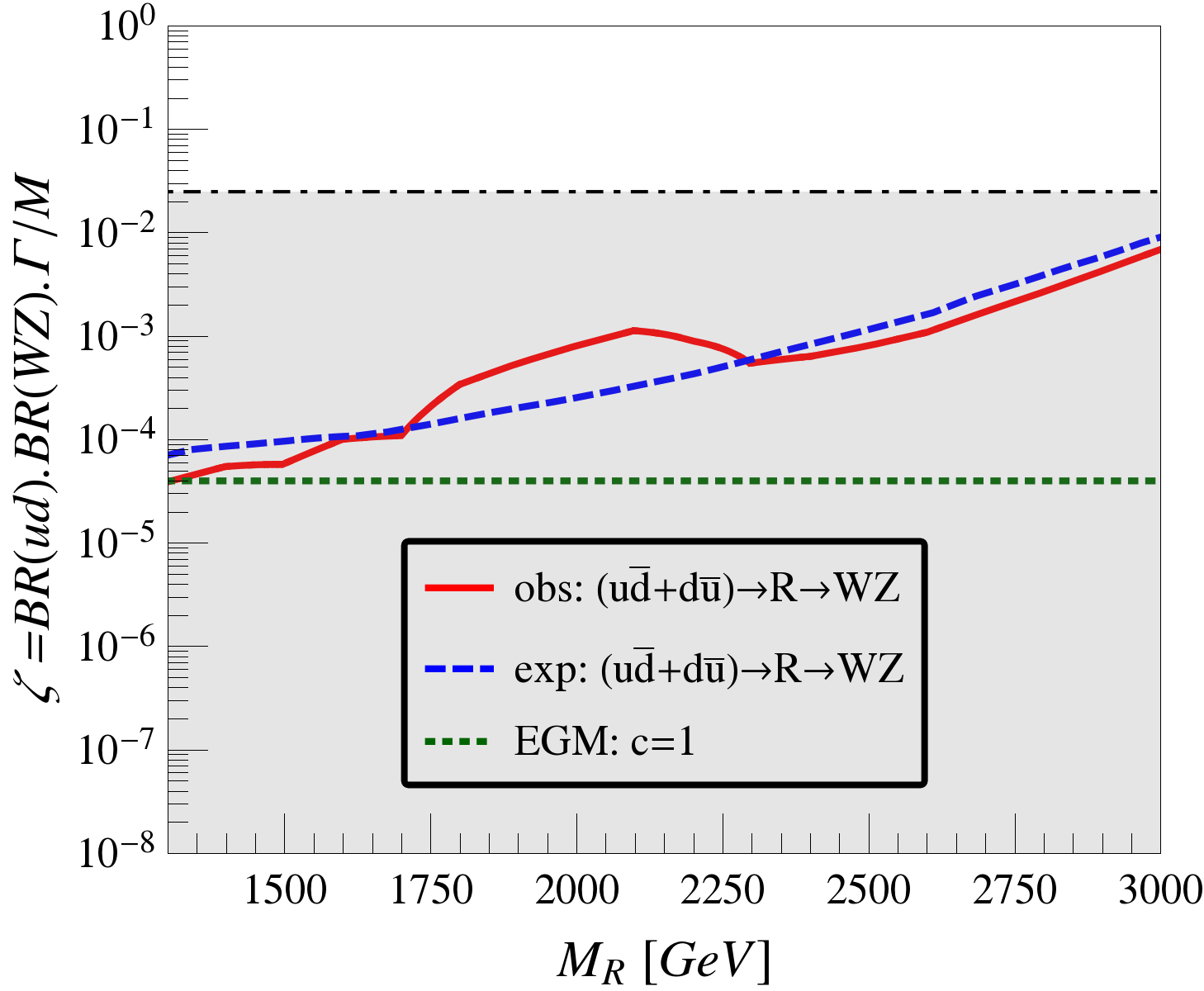}
	\caption{\small \baselineskip=3pt {\textbf{Left: }} The experimental ATLAS \cite{Aad:2015owa} upper limits and expected limits on the production cross-section for $WZ \to R \to WZ$ yield these upper bounds on $\zeta$ assuming production of the s-channel resonance $R$ via vector boson fusion alone. Since the potential excess (the difference between the curves) near 2 TeV corresponds to $\zeta \sim 100$, it lies well outside the allowed (shaded) region. {\textbf{Right: }}The experimental ATLAS \cite{Aad:2015owa} upper limits and expected limits on the production cross-section for $u\bar{d} + d\bar{u}) \to R \to WZ$ yield these upper bounds on $\zeta$ assuming production of the s-channel resonance $R$ via $u\bar{d} + d\bar{u}$ alone. Since the apparent excess corresponds to $\zeta \sim 10^{-4}$, it lies well within the allowed (shaded) region. The extended gauge model (EGM)~\cite{Altarelli:1989ff} with the coupling factor $c$ set to 1 predicts a value of eta well below what would be required to explain the apparent excess.}
	\label{fig:simplified-eta-vbf2-ud2}
\end{figure}

The same comparison is made in the left pane of Figure~\ref{fig:simplified-eta-vbf2-ud2} using the variable $\zeta$ on the vertical axis.  Using $\zeta$ removes the need to show separate curves for different values of $\Gamma/M$.  We see that the value of $\zeta$ corresponding to the possible excess production around 2 TeV would be $\zeta \sim 100$; this is far above the maximum value of 0.1 that forms the upper boundary of the shaded allowed region in the plot; hence, a fermiophobic resonance would not be a viable candidate for producing such an excess.

\subsubsection{Non-fermiophobic $W^\prime$ decaying to dibosons:  $u d \to W^{\prime} \to W^{\pm} Z$}

Next, consider a charged spin-one color-neutral vector resonance -- a technirho or a $W^\prime$ -- that couples to both quark/anti-quark pairs and to vector boson pairs.  Since the parton luminosities for $(c,s,W^{\pm},Z)$ are small, those initial states can be neglected.
Thus, we have a resonance produced via $q\bar{q}$ (in this case primarily $u\bar{d}$ or $d \bar{u}$) annihilation and capable of decaying to vector boson pairs.  Since the 2015 data \cite{Aad:2015owa} showed a potential excess in diboson pairs but not one in dijets, we restrict ourselves to the situation in which the W' couples to $q\bar{q'}$ far more weakly than to $WZ$, so that dijet decays will not be significant.  Therefore, the $W'$ effectively has one production mode and a different single decay mode.

In this case, the signal cross section is determined entirely by $BR(R\to q\bar{q}) \cdot BR(R\to WZ)$, which is bounded from above by $1/4$, since the two incoming partons differ from one another.

In the right pane of Figure~\ref{fig:simplified-vbf2-ud2}, we have applied Eq.~\ref{eq:simplebound-lower} to the observed and expected experimental ATLAS upper limits on the production cross-section \cite{Aad:2015owa} to obtain an upper bound on the product of branching ratios of the resonance into the $ud$ initial state and  $W^\pm Z$ final state. As doing so requires one to assume a value for the resonance's width/mass ratio, we show the results for for $\Gamma_R / M_R = 1\%,\ 10\%$.  In contrast to the results for the fermiophobic $W'$ from the left pane of this figure, here we see that a $W^{\prime}$ produced via $q\bar{q}$ annhilation can be consistent with the observed excesses so long as the corresponding product of the branching ratios to $WZ$ and $q\bar{q}$ lies within the shaded region. 

The same comparison is made in the left pane of Figure~\ref{fig:simplified-eta-vbf2-ud2} using the variable $\zeta$ on the vertical axis.  This removes the need to show separate curves for different values of $\Gamma/M$.  Now we see that the value of $\zeta$ corresponding to the possible excess production around 2 TeV would be $\zeta \sim 10^{-4}$; this is well below the maximum value of 0.025 that forms the upper boundary of the shaded allowed region in the plot, leaving a non-fermiophobic $W^\prime$ as a viable possibility.  The $W^\prime$ boson of the extended gauge model (EGM)~\cite{Altarelli:1989ff} with the coupling factor $c$ set to 1 predicts a value of $\zeta$ well below what would be required to explain the apparent excess.

When comparing the observed upper bound curves (solid red) in the left and right panes of Figure~\ref{fig:simplified-vbf2-ud2} or the left and right panes of Figure~\ref{fig:simplified-eta-vbf2-ud2}, it is clear that they are similar but not identical.  They are similar because they are derived from the same data set, an upper bound on the rate of WZ events at LHC. However, the curves in the left and right panes of these figures are produced under different assumptions about the incoming partons whose fusion produced the $WZ$ final state: incoming $WZ$ for the left panes and incoming quark/anti-quark pairs for the right panes.  Because parton luminosities for $W$ and $Z$ bosons are far smaller than for first generation quarks, the constraints on the product of branching ratios or the quantity $\zeta$ is much more stringent in the right panes.  At the same time, because the PDFs for the light quarks have a different energy dependence than those for the $W$ and $Z$ (the difference varies logarithmically with energy), the dependence on resonance mass of the upper bound curves in the left and right panes is also slightly different.

\subsubsection{Photophillic Resonance: $\gamma \gamma \to R \to \gamma \gamma$}
Let us move on to resonances that may be relevant to the hints of a new diphoton resonance at 750 GeV reported in winter 2015  \cite{ATLAS-Diphoton,Moriond-ATLAS,ATLAS-CONF-2016-018,CMS:2015cwa,CMS:2015dxe,Moriond-CMS,CMS-PAS-EXO-16-018}.  First, we consider a new state (either spin-0 or spin-2) produced by photon fusion and decaying only to diphotons.  Conceptually, this case resembles the fermiophobic $W'$ in that the unique initial and final states are identical; note, however, that the two initial state partons are identical, so that the upper limit on $(1+\delta_{ij})[BR(R\to \gamma\gamma)]^2$ is 2 rather than 1. This example would be of phenomenological interest if a new resonance were seen only in a diphoton decay channel.

\begin{figure}[htbp]
	\centering
	\includegraphics[width=0.49\textwidth]{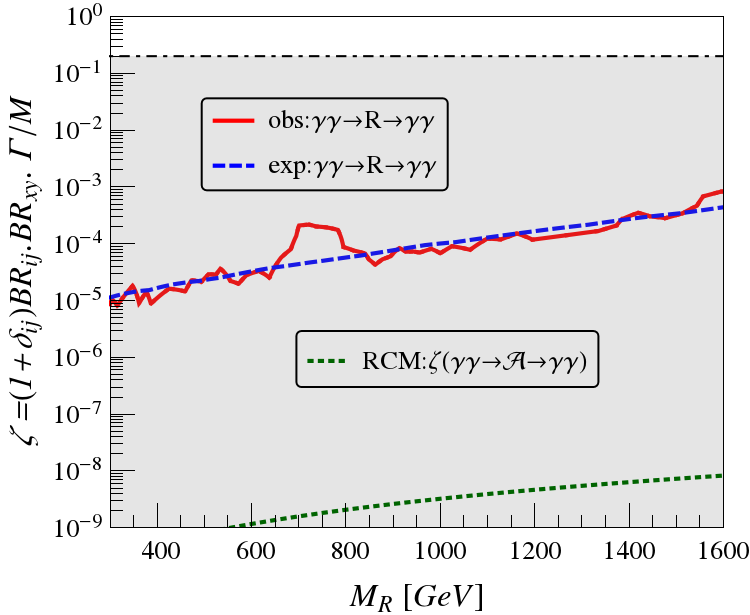}
		\includegraphics[width=0.49\textwidth]{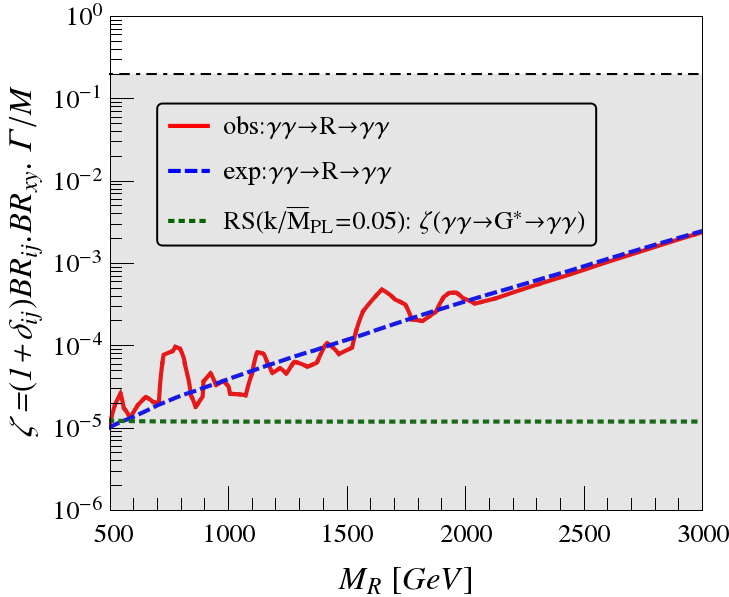}
	\caption{\small \baselineskip=3pt
Experimental observed (solid red) and expected (dashed blue) upper bounds~\cite{Aaboud:2016tru} on $\zeta$ for production of an s-channel  resonance $R$ via photon fusion and subsequent decay to diphotons. Since the contours lie in the (shaded) region where $\zeta$ is below the maximum value for this process, photon fusion alone may be the dominant production mode of such a narrow resonance. \textbf{		Left:} Spin-0 resonance. The green-dotted curve indicates the predicted value of $\zeta$ for the Renormalizable Coloron Model \cite{Chivukula:2016zbe}; it lies several orders of magnitude below the value required to account for the apparent excess events at 750 GeV.  
		\textbf{		Right:} Spin-2 resonance. The green-dotted curve indicates the predicted value of $\zeta$ for the RS Graviton model \cite{Bijnens:2001gh} with the parameter value as indicated; it lies about a factor of five below the value required to account for the apparent excess events at 750 GeV.
	}
	\label{fig:simplified-gaga1}
\end{figure}

The $\gamma \gamma$ luminosities are produced using the {\tt CT14} photon pdfs~\cite{Schmidt:2015zda}. Results are reported in  Figure~\ref{fig:simplified-gaga1}.  The solid curve shows the observed upper bound, while the expected upper bound is denoted by the dashed curve. We will take the difference between the observed and expected upper limits as an indication of the value of $\zeta$ required to produce the excess tentatively seen at a mass of 750 GeV; that points to a value of $\zeta$ of order $10^{-4}$. 

For comparison, the predicted value of $\zeta$ as a function of resonance mass in the Renormalizable Coloron Model (RCM)is shown (dotted green curve), assuming  that the pseudoscalar state is produced via diphoton fusion and decays back to diphotons.\footnote{The values of the parameters of the model are the same as in ref.~\cite{Chivukula:2016zbe}. The number of generations of singlet and doublet vector like quarks are chosen to be $n_q=3$ and $N_Q=3$.} The RCM was proposed in \cite{Bai:2010dj} and has also been studied extensively in \cite{Chivukula:1996yr,Hill:1993hs,Dicus:1994sw,Chivukula:2013xka,Chivukula:2014rka,Chivukula:2015kua,Chivukula:2016zbe}.   Unfortunately, we can see from the figure that the RCM would provide an $\zeta$ value five orders of magnitude too small.  A pseudoscalar produced by photon fusion in the RCM cannot account for the apparent excess.

 The right pane shows a comparison with the theoretically predicted value of $\zeta$ in the RS Graviton model \cite{Randall:1999ee, Bijnens:2001gh} for a spin-2 graviton produced by photon fusion and having $(k/\bar{M}_{Pl} = 0.05$ (dotted green curve). Since the prediction lies a factor of five below the value required to account for the apparent excess events at 750 GeV, it is not obvious that the model, with this choice of parameters, could provide an explanation -- although larger values of $k/\bar{M}_{Pl}$ could potentially accommodate the excess. 

\subsubsection{Boson-Phillic Resonance: $g g \to R \to \gamma \gamma$}

Alternatively, we may consider a (spin-0 or spin-2) resonance that can be produced via gluon fusion and still decays to photons. Conceptually, this resembles the non-fermiophobic $W'$ in that the unique initial and final states differ from one another.  Since the two incoming partons are identical, the upper limit on $(1+\delta_{ij})BR(R\to gg) BR(R\to\gamma\gamma)$ is $1/2$.  Again, this case is of interest if a diphoton resonance is seen without a corresponding dijet signal; the resonance's branching fraction to dijets must be small enough to avoid a dijet signal yet still large enough that gluon fusion is the dominant production mode. 

\begin{figure}[htbp]
	\centering
	\includegraphics[width=0.49\textwidth]{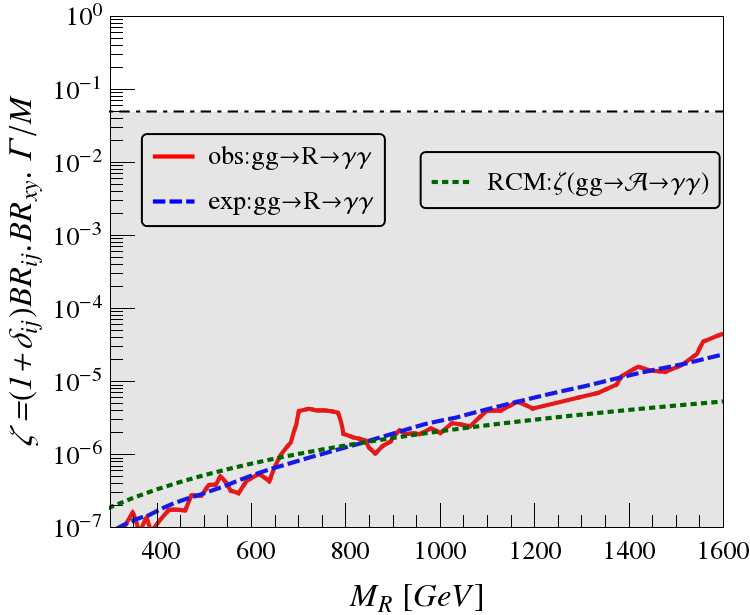}
		\includegraphics[width=0.49\textwidth]{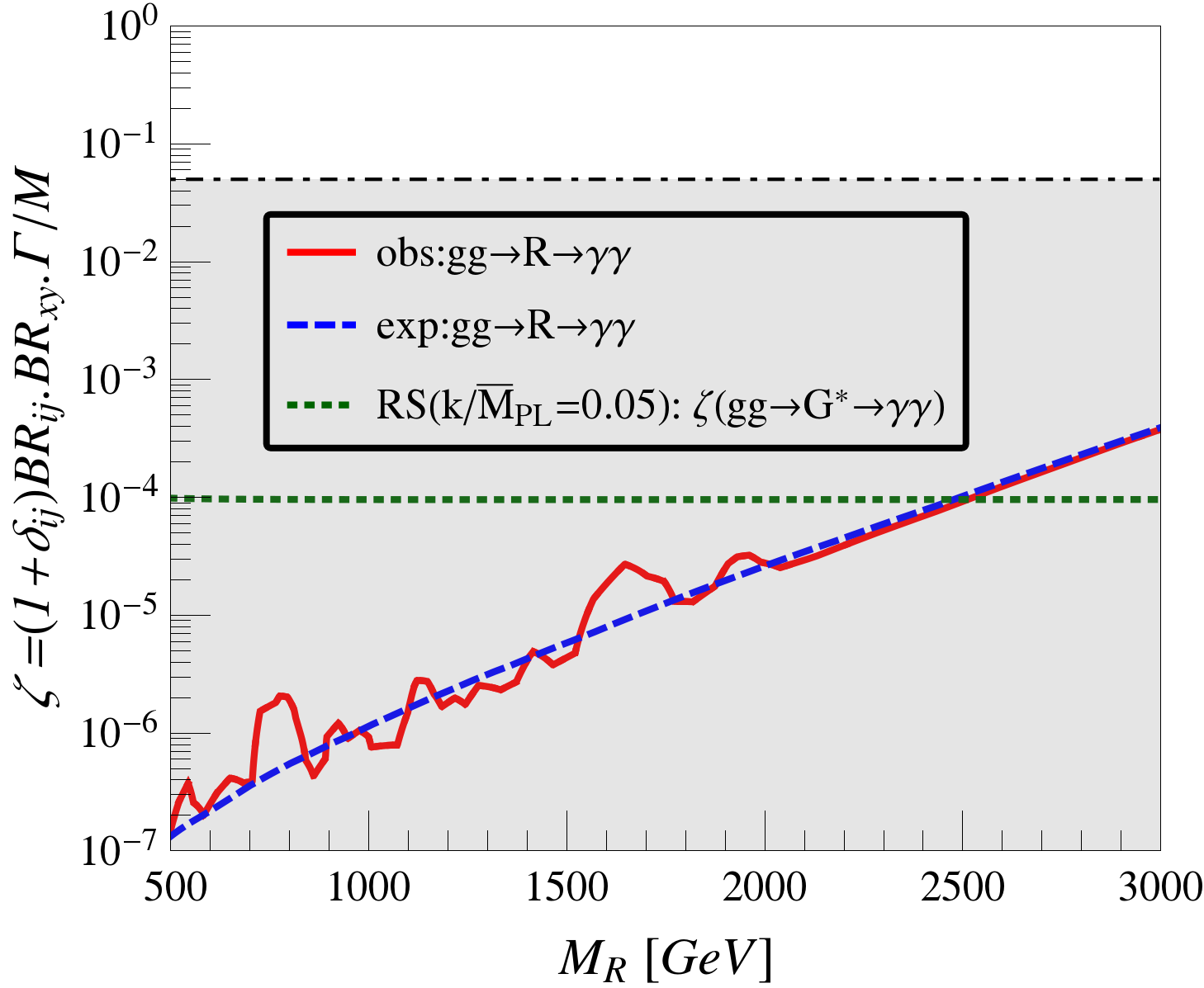}
	\caption{\small \baselineskip=3pt
		Experimental observed (solid red) and expected (dashed blue) upper bounds~\cite{Aaboud:2016tru} on $\zeta$ for production of an s-channel  resonance $R$ via gluon fusion and subsequent decay to diphotons. Since the contours lie in the (shaded) region where $\zeta$ is below the maximum value for this process, gluon fusion alone may be the dominant production mode of such a narrow resonance. \textbf{		Left:} Spin-0 resonance. The green-dotted curve indicates the predicted value of $\zeta$ for the Renormalizable Coloron Model \cite{Chivukula:2016zbe}; it crosses through the window between the observed and expected upper bounds on eta at a resonance mass of order 750 GeV, indicating that the RCM could explain the apparent excess of diphoton events.  
		\textbf{		Right:} Spin-2 resonance. The green-dotted curve indicates the predicted value of $\zeta$ for the RS Graviton model \cite{Bijnens:2001gh} with the parameter value as indicated; it is excluded for resonance masses below about 2.5 TeV, setting a lower bound on the graviton mass.  
	}
	\label{fig:simplified-gg1}
\end{figure}

Results are reported in  Figure~\ref{fig:simplified-gg1}.  The solid (red) curve shows the observed upper bound, while the expected upper bound is denoted by the dashed (blue) curve. Note that the upper bound on $\zeta$ as a function of resonance mass is far more stringent for a resonance produced by gluon fusion (Figure~\ref{fig:simplified-gg1}) than for one produced by photon fusion (Figure~\ref{fig:simplified-gaga1})), because the gluons' parton luminiosity is so much larger.  Similarly, because the PDFs of the gluon and photon have different energy dependences, the slopes of the upper bound curves are also slightly different from one another.
 
 For comparison, in the left pane the predicted value of $\zeta$ as a function of resonance mass in the Renormalizable Coloron Model (RCM)  is shown (dotted green curve), assuming that the pseudoscalar state characteristic of that model is produced via gluon fusion and decays back to diphotons.\footnote{For the RCM, we choose the same parameter values as in ref.~\cite{Chivukula:2016zbe}, and set the number of generation of doublets and singlet to be three each $(N_Q = n_q =3)$.}   If we take the difference between the observed and expected upper limits as an indication of the value of $\zeta$ required to produce the excess tentatively seen at a mass of 750 GeV, that points to a value of $\zeta$ of order $10^{-6}$, which is in line with the RCM prediction. Thus, as discusssed in Ref.~\cite{Chivukula:2016zbe}, the RCM is a viable candidate model to explain such a diphoton excess.   
 
 In the right pane, comparison with the theoretically predicted value of $\zeta$ in the RS Graviton model with $(k/\bar{M}_{Pl} = 0.05$ is shown (dotted green curve).  This illustrates that the RS graviton predicts a value of $\zeta$ that is excluded for resonance masses below about 2.5 TeV, setting a lower bound on the graviton mass. It is therefore not a good candidate to explain a diphoton excess at 750 GeV (the predicted value of $\zeta$ lies several orders of magnitude above the upper bound at that mass). 

\subsection{Examples of the Nearly-Simplest Case: a single production mode, multiple decay modes}

\begin{figure}[htbp]
	\centering
	\includegraphics[width=0.49\textwidth]{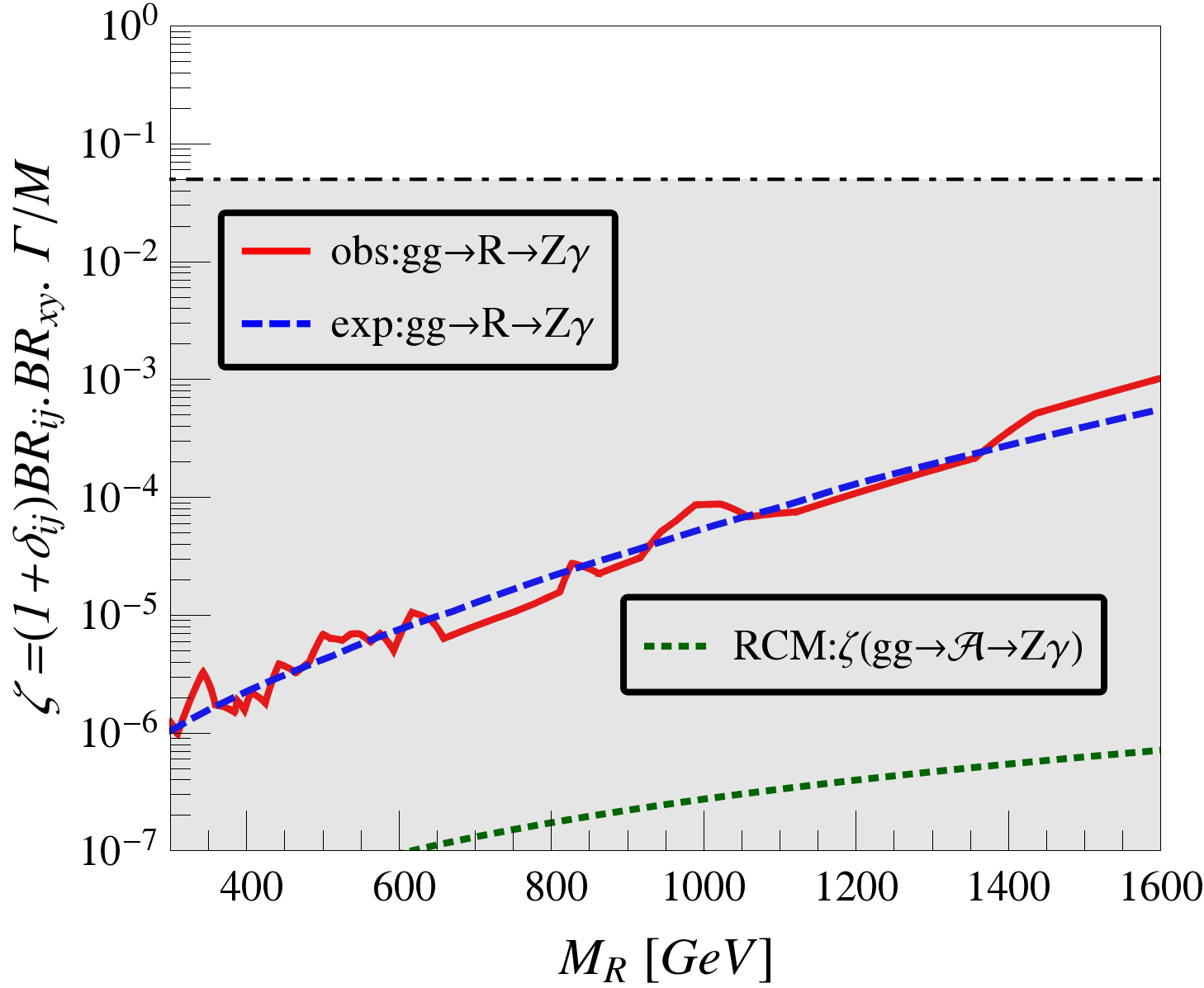}
	\caption{\small \baselineskip=3pt
			Experimental observed (solid red) and expected (dashed blue) upper bounds~\cite{ATLAS-CONF-2016-010} on $\zeta$ for production of an s-channel  resonance $R$ via gluon fusion and subsequent decay to $Z \gamma$. Since the contours lie in the (shaded) region where $\zeta$ is below the maximum value for this process, gluon fusion alone may be the dominant production mode of such a narrow resonance. The green-dotted curve indicates the predicted value of $\zeta$ for the Renormalizable Coloron Model \cite{Chivukula:2016zbe}; it lies well below the observed and expected upper bounds on $\zeta$, indicating that the RCM is not excluded by this data.
			\label{fig:z-gamma}
		}
\end{figure}
Here, at most one of the significant decay modes is available as a production mode because the other decay modes involve states with negligible parton distribution functions. 

One example would be a $W^\prime$ boson with multiple decay modes.  The positively charged state would be produced through $u\bar{d}$ fusion and would decay to dijets or charged lepton plus neutrino or $WZ$.  The product of the production BR and the summed decay BR's would be bounded from above by 1/4 since the incoming partons are not identical.

Another example would be a neutral scalar produced through gluon fusion, which decays both to diphotons and to $Z \gamma$. The product of the production BR and the summed decay BR's would be bounded from above by 1/2 since the incoming partons are identical. Building on the previous analysis of diphoton resonances, we show in Figure~\ref{fig:z-gamma},  limits~\cite{ATLAS-CONF-2016-010} on $\zeta$ that arise due to a scalar resonance produced predominantly through gluon fusion and decaying to a $Z$ boson and a photon. The green dotted curve represents the RCM which lies well below the observed (solid red curve) and expected (blue dashed curve) limits, indicating that the RCM is not excluded by this ATLAS data~\cite{ATLAS-CONF-2016-010}.  At the same time, if any of the small excursions of the observed limit above the expected limit  were taken as a possible indication of an excess, a model purporting to account for that excess would have to produce a $\zeta$ of order $10^{-6}$ to $10^{-4}$; given its small $\zeta$ values, the RCM would not be able to provide an explanation.

\subsection{Examples of the General Case: Multiple production and decay modes }

\subsubsection{$Z^{\prime}$ boson: $u\bar{u}, d\bar{d} \to Z^{\prime} \to jj, b\bar{b}, \ell^+\ell^-$}

Production of a $Z^{\prime}$ boson belongs in this category, since many proposed $Z^{\prime}$ bosons have significant couplings to both up and down quarks, giving $u\bar{u}$ and $d\bar{d}$ annihilation as separate production channels with distinct parton luminosities. 
\begin{itemize}
\item A leptophobic $Z'$ coupling only to light quarks, and being detected through its decays to dijets, would have the product of its summed incoming and summed outgoing branching ratios bounded from above by 1.
\item A leptophobic $Z'$ coupling far more to the third generation than the first and second generations would have the product of [the sum of the dominant incoming BR ($u\bar{u},\, d\bar{d},\, c\bar{c},\, s\bar{s}$)] and [the sum of the dominant outgoing BR ($b\bar{b},\, t\bar{t}$)] bounded from above by 1/4.  However, in practice, searches for a new resonance capable of decaying to both $b\bar{b}$ and $t\bar{t}$ final states are conducted separately in these two channels, since they appear so different in the detector -- and each product of BR's would be bounded by 1/4.
\item If a $Z^{\prime}$ coupling to both quarks and leptons were studied in its decays to charged leptons, the product of the summed (over quarks) production BR's and the summed leptonic decay BR's would be bounded from above by 1/4.
\end{itemize}

\begin{figure}[htbp]
	\centering
	\includegraphics[width=0.48\textwidth]{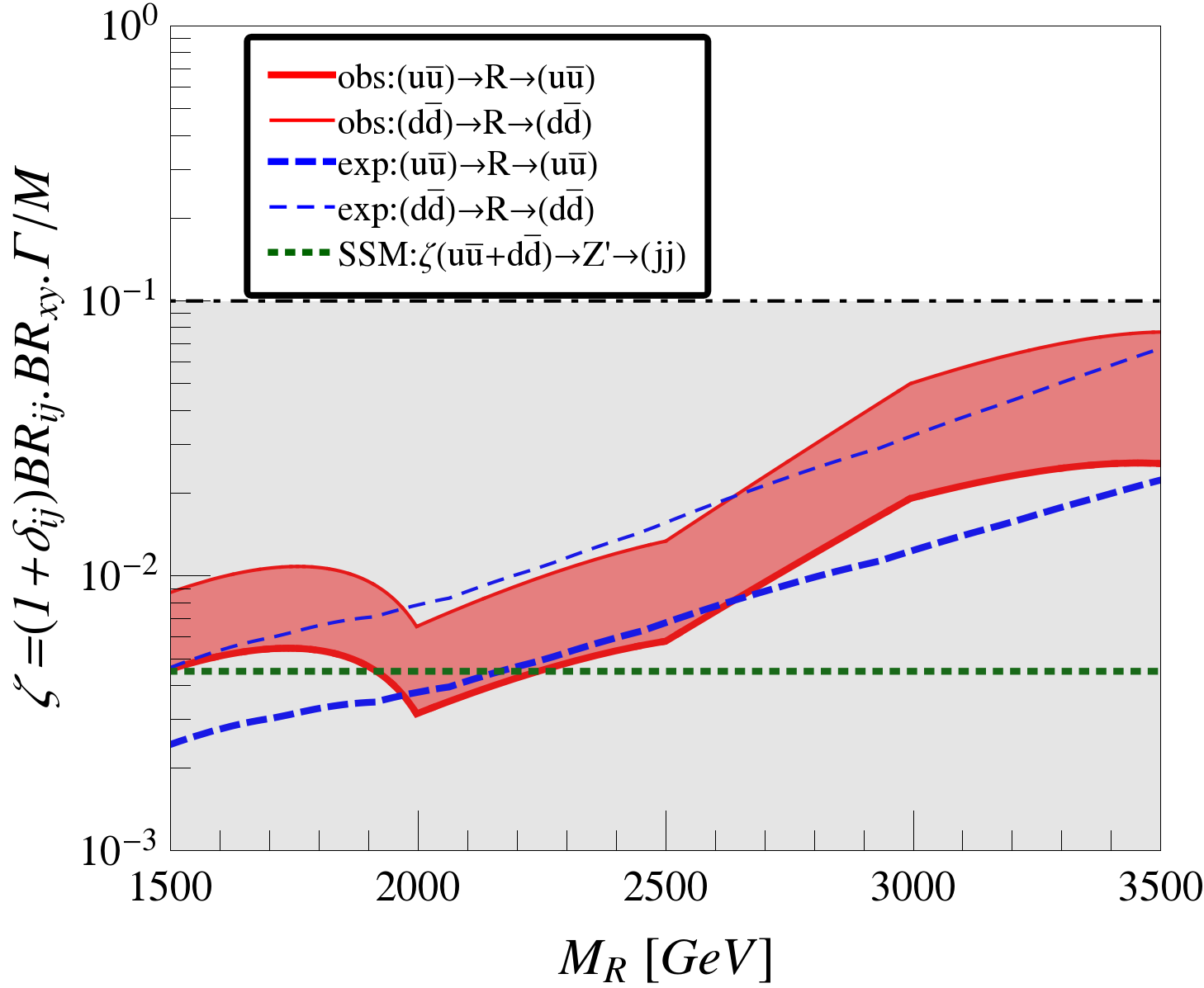}
	\includegraphics[width=0.48\textwidth]{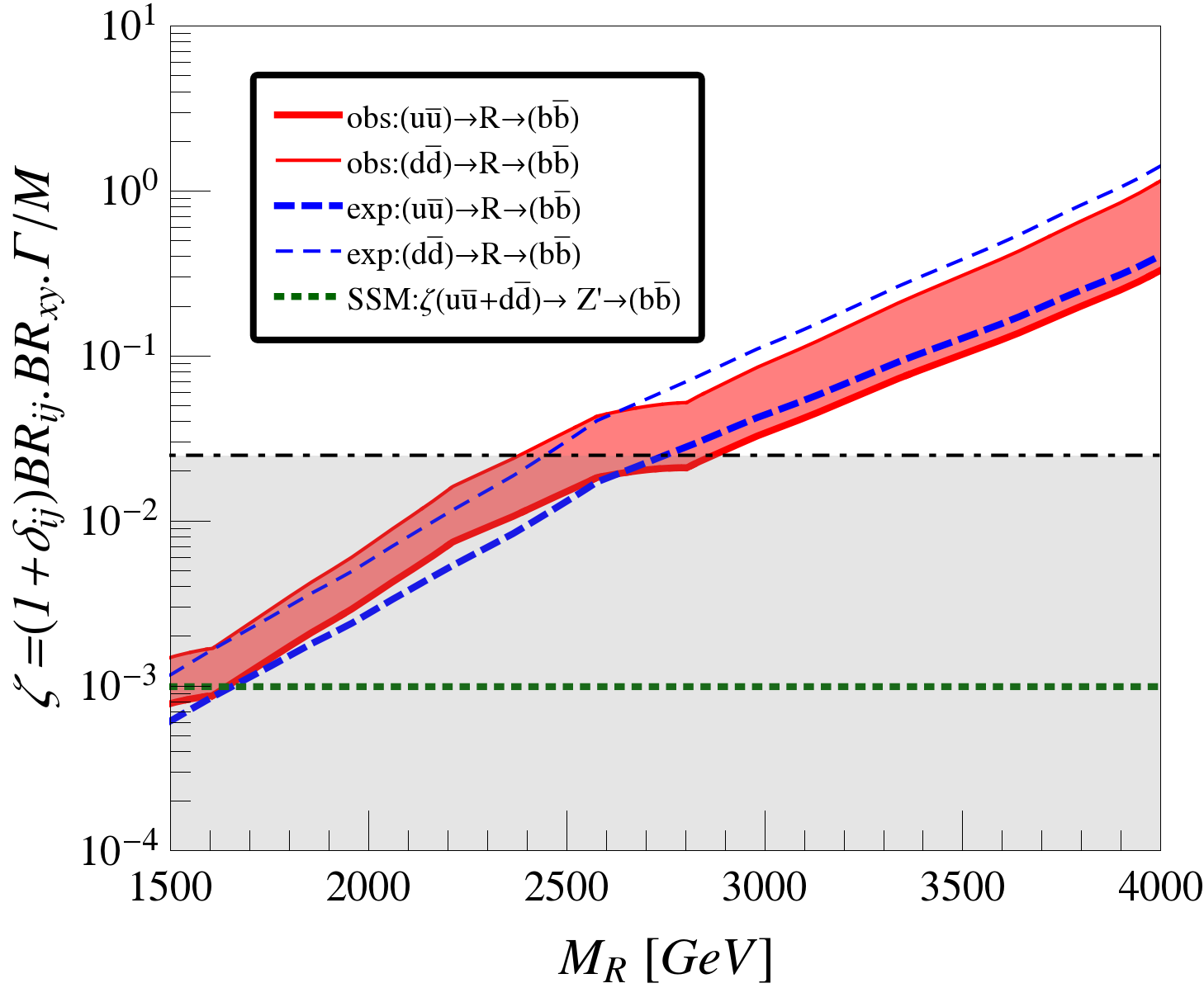}\vspace{0.5cm}
		\includegraphics[width=0.48\textwidth]{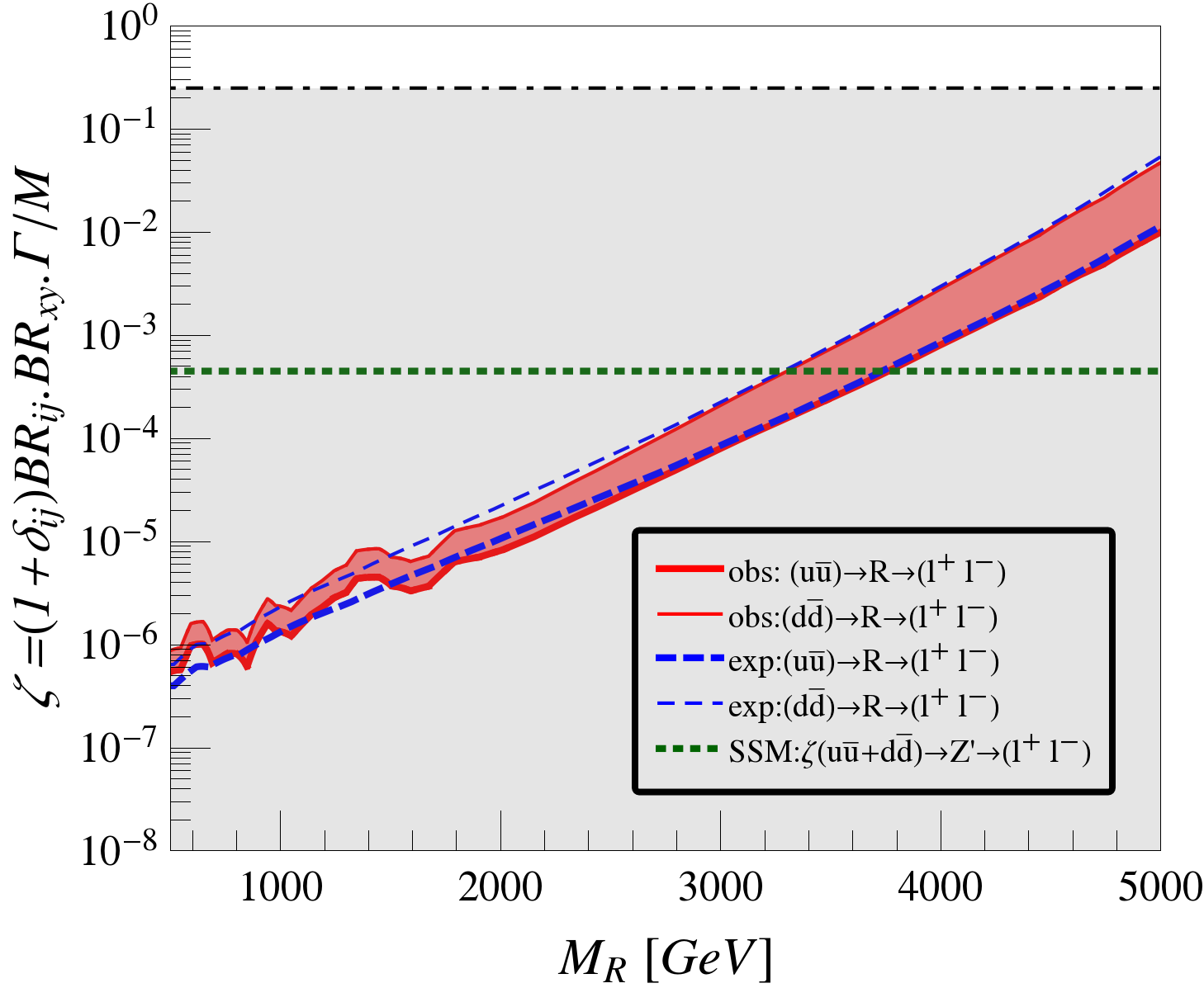}
	\caption{\small \baselineskip=3pt  In these plots in the $\zeta$ vs. resonance mass plane, the shaded region corresponds to values of $\zeta$ consistent with upper bounds on branching ratios as described in the text (1 for the upper left pane and 1/4 for the other panes) and $\Gamma/M \leq 10\%$. The observed (red solid) and expected (blue dashed) upper bounds on $\zeta$ as a function of resonance mass are shown in each pane.  The thick solid red and blue dashed curves correspond to the situation in which the $Z^\prime$ couples only to up-flavor (down-flavor) quarks.  The shaded band between the two red curves represents the range of variation of the upper bound on $\zeta$ as the $Z^\prime$ ranges between the coupling extremes represented by the two red curves.
		\textbf{	Upper	Left:}
				Upper bounds on $\zeta$ for a leptophobic $Z^{\prime}$ produced via light quark/anti-quark annihilation and decaying to dijets (red solid curve) compared with expected upper limit (blue dashes) and the size of $\zeta$ provided by a Sequential Standard Model $Z^\prime$ boson (green dots). If a significant excess were deemed present at masses below 2 TeV, the contribution of this $Z^\prime$ boson would be consistent with it.
		\textbf{	Upper	Right: }Upper bounds on $\zeta$ for a leptophobic $Z^{\prime}$ produced via light quark/anti-quark annihilation and decaying to third generation quarks, compared with expected upper limit (blue dashes) and SSM $Z^\prime$ (green dots). 	\textbf{		Lower: }Upper bounds on $\zeta$ for a  $Z^{\prime}$ produced via light quark/anti-quark annihilation and decaying to dileptons, compared with expected upper limit (blue dashes) and SSM $Z^\prime$ (green dots).	
	}
	\label{fig:simplified-Zp}
\end{figure}

To facilitate the comparison with models, we show $\zeta$ on the vertical axis of the plots showing our results.  Since we are studying narrow resonances, with $\Gamma_R / M_R \leq 10\%$, the upper bound on $\zeta$ in each of the three cases discussed above would be one tenth the bound on the product of branching ratios. The shaded region in each plot of Figure~\ref{fig:simplified-Zp} corresponds to the region obeying that bound in the $\zeta$ vs. resonance mass plane. The observed (red solid) and expected (blue dashed) upper bounds on $\zeta$ as a function of resonance mass are shown in each pane.  The thick (thin) solid red and blue dashed curves correspond to the situation in which the $Z^\prime$ couples only to up-flavor (down-flavor) quarks.  The shaded band between the two red curves represents the range of variation of the observed upper bound on $\zeta$ as the $Z^\prime$ ranges between the coupling extremes represented by the two red curves.  This covers the full range of possibilities for $Z^\prime$ bosons coupling to first-generation quarks.  

The upper left pane of Figure~\ref{fig:simplified-Zp} shows the observed upper bounds on $\zeta$ for a leptophobic $Z^{\prime}$ produced via light quark/anti-quark annihilation and decaying to dijets (red solid) alongside the expected upper limit (blue dashes) \cite{ATLAS:2015nsi}.  The value of  
 $\zeta$ for a Sequential Standard Model (SSM) $Z^\prime$ boson ~\cite{Langacker:2008yv} (green dots) is shown for comparison. 
If one suspected that the difference between the observed and expected upper limit near 1.75 TeV, for instance, corresponded to an excess of events stemming from the presence of a resonance, then the SSM $Z^\prime$ would provide a value of $\zeta$ consistent with that required of the resonance.  However, if one made a similar comparison around 3 TeV, it would be clear that the SSM $Z^\prime$ had too small an $\zeta$ value to be the source of such a postulated excess.
 
The upper right pane of Figure~\ref{fig:simplified-Zp} shows the upper bounds on $\zeta$ for a leptophobic $Z^{\prime}$ produced via light quark/anti-quark annihilation and decaying to b-quarks (red solid) alongside the expected upper limit (blue dashes) \cite{Aaboud:2016nbq} and the value of $\zeta$ for the SSM $Z^\prime$ (green dots).  In this channel, the value of $\zeta$ provided by the model lies well below the current upper limits for resonance masses above about 2 TeV.  Note also that the values of $\zeta$ probed by the data for resonance masses above about 3 TeV lie outside the allowed (shaded) region -- and, hence, if an excess of $b\bar{b}$ events with an invariant mass had been observed in this region, it could not have arisen from a model of this type. We find that the SSM $Z^\prime$ is not bounded by experiments  for the mass range shown in the figure. This is consistent with the results in  Ref.~\cite{Aaboud:2016nbq}.

The lower pane of Figure~\ref{fig:simplified-Zp} shows the upper bounds on $\zeta$ for  a $Z^{\prime}$ produced via light quark/anti-quark annihilation and decaying to dileptons (red solid) alongside the expected upper limit (blue dashes) \cite{ATLAS-CONF-2015-070} and the value of $\zeta$ for the SSM $Z^\prime$ (green dots).  In this channel, the value of $\zeta$ provided by the model is excluded by the data for masses below about 3.5 TeV; this provides a lower bound on the SSM $Z^\prime$ boson mass. A similar bound of about 3.5 TeV, is also obtained for the SSM $Z^\prime$ in Ref.~\cite{ATLAS-CONF-2015-070}.

\section{Discussion}
We are proposing a \sout{model-independent} method for quickly determining whether a small excess observed in collider data could potentially be attributable to the production and decay of a single, relatively narrow, s-channel resonance belonging to a generic category, such as a leptophobic $Z^\prime$ boson or a fermiophobic $W^\prime$ boson. Using a simplifed model of the resonance allows us to convert an estimated signal cross section into general bounds on the product of the dominant branching ratios corresponding to production and decay. This quickly reveals whether a given class of models could possibly produce a signal of the required size at the LHC and circumvents the present need to make laboreous comparisons of many individual theories with the data.  Moreover, the dimensionless variable $\zeta$, which multiplies the product of branching ratios by the width-to-mass ratio of the resonance as defined in Eqn.~\ref{eq:gen-bran-rat}, does an even better job at producing compact and easily interpretable results.

In this work, we began by setting up the general framework for obtaining simplified limits and outlining how it applies for narrow resonances with different numbers of dominant production and decay modes.  We then analyzed applications of current experimental interest, including resonances decaying to dibosons, diphotons, dileptons, or dijets.  In section \ref{subsec:fermiophobic-W-prime} we demonstrated that no fermiophobic $W^\prime$ boson could have conceivably explained the diboson excess spotted in the LHC data in summer 2015.  In contrast, we showed that a generic $W^\prime$ produced through quark/anti-quark annihilation could have readily fitted the bill. We further illustrated how easy it was to compare the calculated value of $\zeta$ for a specific instance of such a $W^\prime$ state (using the left-hand side of Eqn.~\ref{eq:gen-bran-rat}) with the experimental upper bound on $\zeta$ in order to determine whether that particular instance was a viable candidate to explain the excess. While those analyses involved resonances with only a single dominanat production mode and either one or two significant decay modes, the analyses of scalars, our subsequent discussions of gravitons, and $Z^\prime$ bosons demonstrated how readily the simplified limits method may be used in more general cases with multiple significant production and decay modes.

If the LHC experiments report their searches for resonances beyond the standard model in terms of the simplified limits variable $\zeta$, alongside the commonly used $\sigma \cdot BR$ now employed, this would make it far easier and swifter for the community to discern what sorts of BSM physics might underly any observed deviations from SM predictions. Instead of sifting through very specific theories one by one, we could first winnow the general classes of resonances and pursue only models incorporating the viable classes of resonances.  Moreover, for a given model, it is straightforward to calculate the value of $\zeta$ for the candidate resonances and do a quick comparison with the data.

Since situations may well arise where new physics makes its first appearance as a scattering excess that is not obviously due to a narrow s-channel resonance that can be fully reconstructed, we are presently extending our analysis.  We are generalizing our simplified limits framework to handle resonances of moderate width treated in the Breit-Wigner approximation and cases where invisible final-state particles necessitate the use of a transverse mass variable (rather than invariant mass).  We are also exploring the boundary between describing the transtion between the initial partons and final state particles in terms of an unresolved four-body contact interaction and in terms of an intermediate resonant state.  These results will be presented in forthcoming work.

\section*{Acknowledgments}

 The work of. R.S.C., K.M., and  E.H.S. was supported by the National Science Foundation under Grants PHY-0854889 and PHY-1519045.  R.S.C. and E.H.S. also acknowledge the support of NSF Grant PHYS-1066293 and the hospitality of the Aspen Center for Physics during work on this paper. P.I. is supported by ``CUniverse" research promotion project by Chulalongkorn University (grant reference CUAASC).

\appendix

\section{Impact of Resonance Width on the Simplified Limits}
\label{app:width}

\begin{figure}
	\centering
	\includegraphics[width=0.5\textwidth]{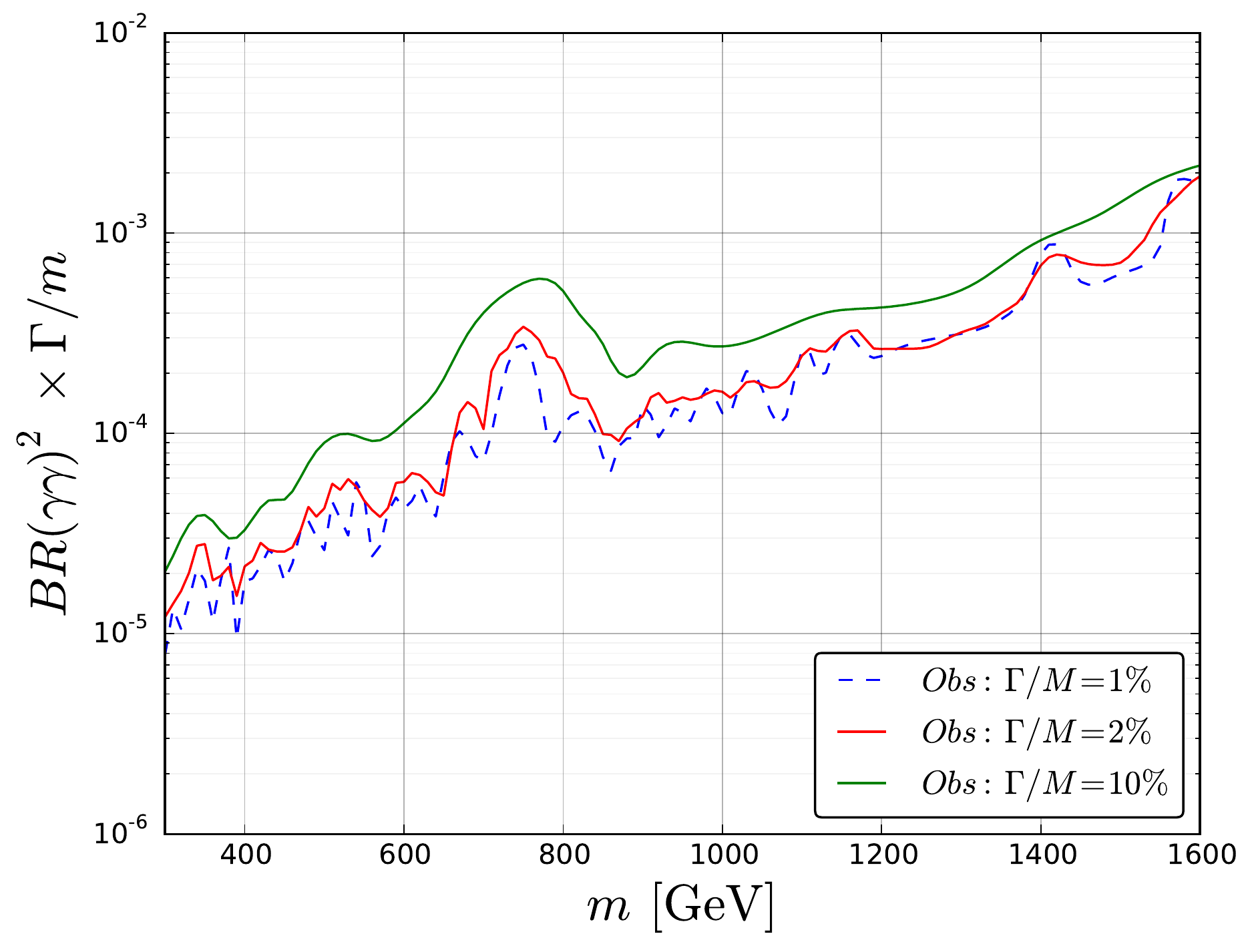}
	\caption{Variation with resonance width of the observed upper bound on $\zeta$ for a scalar produced by photon fusion and decaying to diphotons, across a range of scalar masses; based on data in \cite{Aaboud:2016tru}. The upper limit on $\zeta$ for a scalar with $\Gamma/M = 1\%$ (blue dashed curve) is about a factor of two stronger than that for a scalar with $\Gamma/M = 10\%$ (upper (green) solid curve). }
		\label{fig:width-variation-limits}
\end{figure}

We have worked in the narrow width approximation in establishing limits on the value of $\zeta$.  This encompasses widths ranging from well below the detector resolution up to about ten percent of the resonance's mass.  The experimental limits on production cross-sections do depend on the precise width assumed for one of these narrow resonances.  One may see this by comparing, for instance, the limits that ATLAS has established on a diphoton resonance in \cite{Aaboud:2016tru}; the limits on a narrow (4 MeV width) resonance are clearly stronger than those on a resonance assumed to have $\Gamma/M = 10\%$.  As illustrated in Figure~\ref{fig:width-variation-limits}, we estimate that for scalars produced via photon fusion and decaying to diphotons, the upper limit on $\zeta$ for a resonance with a width equal to 1\% of the mass is about a factor of two stronger than that for a resonance of 10\% width, across the range of masses included in our analysis. This variation reflects the precision of the simplified limits discussed here.

\section{Parton Luminosity}
\label{app:parton-lum}
The first step in calculating the cross-section given in Eqn.~\ref{eq:simplest2}, is to evaluate parton luminosities.
For this purpose we use {\tt LHAPDF6}~\cite{Buckley:2014ana} as the interface to various parton distribution functions. 

\subsection{Quark and Gluon luminosities}
When the two incoming partons at a p-p collider are some combination of quarks or gluons, the parton luminosity is defined as follows:
\begin{equation}
\frac{dL_{ij}}{d\hat{s}}=\frac{1}{s}\int^{1}_{\epsilon=\hat{s}/s} \frac{1}{1+\delta_{ij}}
\left[
(f_i(x_1,\hat{s})\times f_j(\epsilon/x_1,\hat{s})) + (i \longleftrightarrow j)
\right]
\frac{dx_1}{x_1}~.
\end{equation}
Here $x_1$ is the parton momentum fraction, $\hat{s}= m_R^2$ and $s$ is the square of the energy of the two colliding protons.

\subsection{$W$ and $Z$ boson luminosities}
The $W$ and $Z$ boson parton luminosities can be calculated using the Effective $W$ Approximation~\cite{Dawson:1984gx,Cahn:1983ip,Dawson:1984gc, Kane:1984bb}.
However a few limitations of this approach should be considered carefully.
In this approximation, transverse and longitudinally polarized gauge bosons are considered separately and it is not possible to calculate the contribution to the cross-section from the interference between the longitudinal and transverse component. 
Therefore the effective $W$ approximation is useful only when the amplitude of a process is dominated either by the longitudinal or transverse component.

Furthermore, the effective $W$ approximation requires (a) taking the intermediate vector bosons on-shell and (b) that the intermediate bosons are produced at small angles to the incoming quarks and (c) that the momentum fraction $x$ of either incoming parton satisfy
\begin{equation}
x\gg \frac{2 M_V}{\sqrt{\hat{s}}}.
\end{equation}
For resonant production of a particle with mass $M$, this last constraint translates to $M \gg 2 M_V$.

\subsubsection{Transverse gauge bosons}
In the high energy limit, when the quark energy $E \gg M_v$ the probability distribution of transverse $W$'s or $Z$'s in a quark can be written down as~\cite{Dawson:1984gx}
\begin{equation}
f_{q/V^t}(x) \simeq 
\frac{C_v^2 + C_a^2}{8 \pi^2 x}
\left(
x^2 + 2(1-x)
\right)
\text{log}\left(\frac{4E^2}{M_V^2}\right).
\label{eq:pdfVt}
\end{equation}

For $V=W$
\begin{gather}
C_v=C_a = \frac{g_{2}}{\sqrt{2}}, \\ \nonumber
g_{2}=e/\sin\theta_W
\end{gather}
and for $V=Z$
\begin{gather}
C_v= \frac{g_{2}}{\cos\theta_W}
\left(
\frac{1}{2}T_{3L} -Q \sin^2\theta_W 
\right), \\ \nonumber
C_a= \frac{g_{2}}{\cos\theta_W}
\left(
\frac{1}{2}T_{3L}  
\right).
\end{gather}
$T_{3L}$ is the third component of weak isospin of the quark off which the $V$-boson is being radiated and $Q$ is its electric charge.

To obtain the distribution of vector-bosons inside the proton $(f_{p/V^t}(x))$, one must fold the pdf in Eqn.~\ref{eq:pdfVt} with the quark pdf $(f_{i}(x))$.
\begin{equation}
f_{p/V^t}(x)=
\sum_{i}\int_{x}^{1}
\frac{dx_1}{x_1}
f_{i}(x_1)f_{q_i/V^t}\left(\frac{x}{x_1}\right)\ ,
\label{eq:pdf-pVt}
\end{equation}
where the sum over $i$ runs over all relevant partons.
The luminosity for the two intermediate vector bosons can be found using 
\begin{eqnarray}
\frac{dL_{q_i q_j/V^tV^t}}{d\tau}&=& 
\int_{\tau}^{1}
\frac{dx}{x} (
f_{q_i/V^t}(x) f_{q_j/V^t}(\tau/x)
\\
&=&
\left(\frac{C_{vi}^2 + C_{ai}^2 }{8 \pi^2}\right)
\left(\frac{C_{vj}^2 + C_{aj}^2 }{8 \pi^2}\right)
\frac{1}{\tau}\\
&\times&
\left[
(2+\tau)^2\log(1/\tau) - 2(1-\tau)(3+\tau)
\right]
\log\left(\frac{4E_i^2}{M_V^2}\right)
\log\left(\frac{4E_j^2}{M_V^2}\right)\ .
\end{eqnarray}
Convoluting the above result with the quark pdfs gives the vector boson luminosity in proton-proton collisions:
\begin{equation} 
\frac{dL_{pp/V^tV^t}}{d\tau}=
\sum_{i,j}
\frac{1}{1 +\delta_{ij}}
\int_{\tau}^{1}\frac{dx_1}{x_1}
\int_{\tau/x_1}^{1}\frac{dx_2}{x_2}
\left(
f_{i}(x_1)f_{j}(x_2)\frac{dL_{q_iq_j/V^tV^t}}{d\hat{\tau}}
+ i \longleftrightarrow j
\right),
\label{eq:lumppVVt}
\end{equation}
where $\hat{\tau}=\tau/(x_1 x_2)$.
This can be derived from 

\begin{eqnarray} 
\frac{dL_{pp/V^tV^t}}{d\tau}&=&
\int dx_1 dx_2
\int dy_1 dy_2
\left(
f_{i}(x_1)f_{j}(x_2)
f_{q_i/V^t}(y_1) f_{q_j/V^t}(y_2)
\delta(x_1y_1x_2y_2 - \tau)
\right) \nonumber \\
&=&
\int
\frac{dx_1 dx_2}{x_1 x_2}
f_{i}(x_1)f_{j}(x_2)
\int_{\hat{\tau}}^{1}
\frac{dy_1}{y_1} 
\left(
f_{q_i/V^t}(y_1) f_{q_j/V^t}(\hat{\tau}/y_1) \ .
\right) \nonumber
\end{eqnarray}
Alternatively the expression in Eqn.~\ref{eq:lumppVVt} can be derived by convoluting the pdfs defined in Eqn.~\ref{eq:pdf-pVt}.
\begin{equation}
\frac{dL_{pp/V^tV^t}}{d\tau}=
\int_{\tau}^{1} \frac{dx}{x}
f_{p/V^{t}}(x) f_{p/V^{t}}(\tau/x).
\end{equation}

\subsubsection{Longitudinal Gauge Bosons}

The probability distribution of Longitudinal gauge bosons in a quark (when the energy of the quark $E \gg M_V$) can be written as follows
\begin{equation}
f_{q/V^t}(x) \simeq 
\frac{C_v^2 + C_a^2}{4 \pi^2 x}
\frac{1-x}{x}\ .
\label{eq:pdfVL}
\end{equation}
The luminosity for the two intermediate vector bosons is given by
\begin{eqnarray}
\frac{dL_{q_i q_j/V^LV^L}}{d\tau}&=& 
\int_{\tau}^{1}
\frac{dx}{x} \left(
f_{q_i/V^L}(x) f_{q_j/V^L}(\tau/x)
\right)
\\
&=&
\left(\frac{C_{vi}^2 + C_{ai}^2 }{4 \pi^2}\right)
\left(\frac{C_{vj}^2 + C_{aj}^2 }{4 \pi^2}\right)
\frac{1}{\tau}
\left[
(1+\tau)\log(1/\tau) - 2(1-\tau)
\right].
\end{eqnarray}
Note that unlike the transverse polarization luminosity, the longitudinal luminosity (in the large energy limit) is independent of the energy of the quarks.

\bibliography{simlim_refs-updated}

\end{document}